\documentclass[preprint,showpacs]{revtex4}
\usepackage[centertags]{amsmath}
\usepackage{amsfonts}
\usepackage{amssymb}
\usepackage{amsthm} 
\usepackage{newlfont}
\usepackage{epsfig}
\usepackage{amscd}
\usepackage{graphicx}
\usepackage{epsfig}
\usepackage{amscd}
\usepackage{color}

\newcommand{\beq}{\begin{equation}}
\newcommand{\eeq}{\end{equation}}
\newcommand{\ba}{\begin{array}}
\newcommand{\ea}{\end{array}}
\newcommand{\bea}{\begin{eqnarray}}
\newcommand{\eea}{\end{eqnarray}}
\begin{document}

\begin{center}
{\large \sc \bf {
Quantum correlations responsible for remote state creation:
strong and weak control parameters.
}}

\vskip 15pt

{\large 
S.I.Doronin and A.I.~Zenchuk 
}

\vskip 8pt

{\it Institute of Problems of Chemical Physics, RAS,
Chernogolovka, Moscow reg., 142432, Russia},\\

\end{center}


\begin{abstract}

We study the  quantum correlations between the two remote qubits (sender and receiver) 
connected by the transmission line (homogeneous spin-1/2 chain)  depending  on 
the parameters of the sender's and receiver's initial states (control parameters). 
 We consider two different measures of quantum correlations: 
the entanglement (a traditional measure) and the informational correlation (based on the parameter  exchange  between the sender and receiver).
 We find the domain in the control parameter space yielding (i) zero entanglement between the sender and receiver during the whole
evolution period and (ii) non-vanishing  informational correlation between the sender and receiver, thus showing that the informational correlation 
is responsible for the remote state creation. 
Among the control parameters, there are  the strong parameters (which strongly 
effect the values of studied measures) 
and the weak ones (whose effect  is negligible), therewith the eigenvalues 
of the initial state are given a
 privileged role.
We also  show that the problem of  small entanglement (concurrence) in quantum information processing 
is similar (in certain sense)  to the problem of small determinants in 
linear  algebra. A particular model of  40-node spin-1/2 communication line is presented.
 \end{abstract}

\maketitle

\section{Introduction}
\label{Sec:Introduction}

The formation and  evolution of quantum correlations  is one of the central problems
of quantum information.  Although 
quantum correlations are necessary to provide advantages of quantum information devices in comparison with their
classical counterparts, { the appropriate   measure of these correlations is not well-established yet. For a long time, quantum entanglement
 \cite{HW,Wootters}
was considered as a suitable measure, but recently 
quantum non-locality \cite{BDFMRSSW,HHHOSS,NC} and speedup 
\cite{Meyer,DSC,DFC,DV,LBAW} were observed in systems with 
minor entanglement. Therefore, the quantum discord was introduces as an alternative measure \cite{Z0,HV,OZ,Z}. 
Still it is not  clear whether the above mentioned  quantum entanglement (even if its  value is minor) 
captures all those quantum correlations that  provide the advantages of any quantum device, or other types of correlations
(which are  captured, for instance, by discord rather then by   entanglement) become more important in certain cases.}

{ We may assume  that the quantum correlations can be classified  (with possible overlaps among different classes)
so that a given quantum process is governed by a certain class of quantum correlations rather then by all of them.}
In this paper we are aimed on revealing  those  quantum correlations that are responsible  for remote state creation 
\cite{ZZHE,BPMEWZ,BBMHP,PBGWK2,PBGWK,DLMRKBPVZBW,XLYG,LH},
which is the further development of  the problem of  end-to-end quantum  state 
transfer along a spin chain
\cite{Bose,CDEL,ACDE,KS,GKMT,WLKGGB,NJ,SAOZ,CS,BOWB,ABCVV}.  

{
We consider  a model of 
two remote one-qubit subsystems (called the sender 
($S$) and the receiver ($R$)) connected  to each other through the homogeneous spin-1/2 chain (called 
the transmission line ($TL$)). 
The initial state of the whole system is separated one with the both sender and receiver are in  mixed states. 
Therefore, there is no quantum correlations between
the sender and receiver initially. 
After the initial state is installed,  
the state of the whole system evolves under some Hamiltonian giving rise to  mutual quantum correlations 
between the sender and receiver. 

{ 
Our study is based on the comparative analysis of two classes of quantum correlations. The first class is captured 
by the sender-receiver entanglement} {(SR-entanglement)}, while the second class is captured by the 
so-called informational correlation 
 which  has been recently introduced \cite{Z_inf0,Z_inf}.
The latter quantity  counts the number of parameters of the  local unitary
transformation initially applied to the sender (which are called  eigenvector control-parameters below) that can 
be detected at the receiver. 
 Informational correlation  is discrete by its definition and it is
directly related to the associated system of   linear algebraic 
equations whose  solvability turns into the appropriate determinant condition \cite{Z_inf}. 
}

We study the dependence of  { both the SR-entanglement and informational correlation}  on the parameters of the sender's and receiver's initial state 
which we call the control parameters naturally separated  into eigenvalue and eigenvector parameters.
In turn, we separate the eigenvector control-parameters into the 
strong  control parameters 
(whose  values  strongly effect the quantum correlations) and weak control parameters  (whose effect is 
negligible).

In addition, our study  shows that there is a domain in the control parameters space which yields zero SR-entanglement  during, at least,  the considered 
evolution period. However, the parameters from this domain can also be transfered from  the sender to  the receiver (or vise-versa); 
therefore the informational correlation is non-zero and remote state creation is possible. Thus, the informational correlation serves as a measure selecting the 
quantum correlations responsible for the state transfer/creation, { while the SR-entanglement doesn't capture the required correlations}.

At last, we show that the  states with small SR-entanglement have the same  pre-image  in the control-parameter space as 
the states with small determinants.
Therefore, the case of  small determinants  is likely to be 
the case when the advantage of  quantumness disappears. 
In addition, this situation has 
something in common with so called fluctuations of 
entanglement \cite{FY,Y} showing that the value of these fluctuations can reach
the value of entanglement itself, { so that the calculated value of entanglement is 
 not reliable  in that case. }

The paper is organized as follows.
In Sec.\ref{Section:model}, we discuss the initial state of the communication line and classify the control parameters associated with this  state.
The SR-entanglement (SR-concurrence) and informational correlation as two different measures of quantum correlations 
 are discussed in Sec.\ref{Section:measures}. 
 A  particular model of quantum communication line  based on the nearest-neighbor XY Hamiltonian is considered in Sec.\ref{Section:example}.
 Both sender and receiver are one-qubit subsystems in our case. The brief comparative analysis 
  of SR-concurrence and informational correlation is represented in Sec.\ref{Section:entinf}. Finally, the
basic results are discussed in Sec.\ref{Section:conclusion}.
Some additional details concerning the permanent characteristics of communication line, explicit form of the receiver's density matrix, properties of determinants,  the time instant for state registration    
are given in Appendix, Sec.\ref{Section:appendix}.

\section{Classification of control parameters
}
\label{Section:model}
\subsection{Initial state}
We consider a homogeneous spin-1/2 chain  whose evolution is governed by 
some Hamiltonian commuting with the $z$-projection of the total spin momentum
(the external magnetic field is $z$-directed). 
The whole $N$-spin communication line consists of three interacting subsystems: the one-qubit sender $S$ (the first qubit of the chain), 
the one-qubit receiver $R$ (the last qubit of the chain) and the transmission line $TL$ (a spin-chain connecting the sender and receiver). For the sake of simplicity, we consider the tensor-product  initial state
\begin{eqnarray}\label{inst}
\rho_0 = \rho^S_0 \otimes \rho^{TL}_0\otimes \rho^{R}_0,
\end{eqnarray}
 where   $ \rho^S_0$, $ \rho^{TL}_0 $ and $\rho^S_0$ are, respectively, 
 the  initial density matrices of the sender, transmission line and receiver,
 therewith $\rho^{TL}_0$ is the density matrix of the ground state,
 \begin{eqnarray}\label{TL}
 \rho^{TL}={\mbox{diag}}(1,0,0,\dots),
\end{eqnarray}
 and 
\begin{eqnarray}\label{inSR}
\rho^S_0 = U^S \Lambda^S (U^S)^+,\;\; \rho^R_0 = U^R \Lambda^R (U^R)^+.
\end{eqnarray}
Here the eigenvalue and eigenvector matrices read, respectively,
\begin{eqnarray}\label{ev}
\Lambda^S={\mbox{diag}}(\lambda^S,1-\lambda^S),\;\;\;\Lambda^R={\mbox{diag}}(\lambda^R,1-\lambda^R),
\end{eqnarray}
and
\begin{eqnarray}\label{US}
U^S=\left(
\begin{array}{cc}
\cos \frac{\pi \alpha_1}{2} & - e^{-2 i \pi \alpha_2} \sin \frac{\pi \alpha_1}{2}\cr
e^{2 i \pi \alpha_2} \sin \frac{\pi \alpha_1}{2} & \cos \frac{\pi \alpha_1}{2} 
\end{array}
\right),
\end{eqnarray}
\begin{eqnarray}\label{UR}
U^R=\left(
\begin{array}{cc}
\cos \frac{\pi \beta_1}{2} & - e^{-2 i \pi \beta_2} \sin \frac{\pi \beta_1}{2}\cr
e^{2 i \pi \beta_2} \sin \frac{\pi \beta_1}{2} & \cos \frac{\pi \beta_1}{2} 
\end{array}
\right).
\end{eqnarray}
The parameters $\lambda^S$, $\lambda^R$ are referred to as the  eigenvalue control-parameters, while 
the parameters  $\alpha_i$, $\beta_i$ ($i=1,2$) are called the
eigenvector control-parameters.
Their  variation intervals  are following:
\begin{eqnarray}
\label{intervals}
\begin{array}{ll}
0\le \alpha_i \le 1,& 0\le \beta_i \le 1,\;i=1,2,\cr
0\le \lambda^S \le 1, & 0\le \lambda^R \le 1.
\end{array}
\end{eqnarray}
Studying the correlations between the sender and the receiver we need the density 
matrix of the subsystem $SR$, which reads
\begin{eqnarray}\label{RhoSR}
\rho^{SR}(t) ={\mbox{Tr}}_{TL} \Big(V(t) \rho_0 V^+(t) \Big),\;\;\; V(t)=e^{-i H t},
\end{eqnarray}
where the trace is taken over the transmission line.

\subsection{Three types of control parameters
}
\label{Section:cc}

The control parameters $\lambda^S,\;\lambda^R$, $\alpha_i$, $\beta_i$ ($i=1,2$) introduced in formulas (\ref{ev}-\ref{UR}) 
can be separated into three following  groups.
\begin{enumerate}
\item
The two  eigenvalue parameters    $\lambda^S$ and $\lambda^R$.
\item
The  two  parameters $\alpha_1$ and $\beta_1$, characterizing the absolute values of the 
independent eigenvector components of the sender's and receiver's initial states (they are called strong parameters in Sec.\ref{Section:angles}). 
\item 
The two phase-parameters
$\alpha_2$ and $\beta_2$ of the sender's and receiver's initial states (they are called weak parameters in Sec.\ref{Section:angles}).
\end{enumerate}

Studying the 
effects of different parameters 
on the measure of quantum correlations $F$ (either SR-entanglement or informational correlation), we, first of all, consider the mean value, 
{ $\bar{F}$},  of this quantity  with respect to 
 all the eigenvector parameters 
$\tilde \Gamma$,
\begin{eqnarray}
\tilde\Gamma=\{\alpha_1,\alpha_2,\beta_1,\beta_2\},
\end{eqnarray}
selecting the eigenvalues $\lambda^S$ and $\lambda^R$ as the most important control parameters.
The  mean value of a  function $F$ with respect to some  parameter $\gamma$ is defined as follows: 
\begin{eqnarray}\label{avf}
\left\langle
F 
\right\rangle_\gamma = 
\int_0^1 d\gamma   F(\gamma),
\end{eqnarray}
where we take into account that 
 all the  parameters $\alpha_i$ and  
$\beta_i$ have the same variation interval from 0 to 1, as given in (\ref{intervals}).
Thus, the resulting mean value { $\bar{F}$} as a function of the eigenvalues $\lambda^S$ and $\lambda^R$ reads:
\begin{eqnarray}\label{avF}
\bar F(\lambda^S,\lambda^R) = \left\langle 
F 
\right\rangle_{\tilde \Gamma} \equiv \left\langle\left\langle\left\langle\left\langle F
\right\rangle_{\alpha_1}\right\rangle_{\beta_1}\right\rangle_{\alpha_2}
\right\rangle_{\beta_2}.
\end{eqnarray}
Next, to estimate the effect of different eigenvector parameters, we  introduce the so-called standard deviation
with respect to the particular parameter $\gamma\in \tilde \Gamma$. This deviation is also 
a function of the eigenvalues $\lambda^S$ and $\lambda^R$:
\begin{eqnarray}\label{delta}
\delta^{(F)}_\gamma(\lambda^S,\lambda^R)  = 
\sqrt{\Big\langle \Big(\bar F(\lambda^S,\lambda^R)  - 
\langle F \rangle_{\tilde\Gamma_\gamma}\Big)^2 \Big\rangle_\gamma},
\end{eqnarray}
where $\tilde\Gamma_\gamma$ is the list $\tilde\Gamma$ 
without the  parameter $\gamma$, for instance $\tilde \Gamma_{\alpha_1}=
\{\alpha_2,\beta_1,\beta_2\}$.

The two measures of quantum correlations (denoted by $F$ in the above formulas)  are briefly described in Sec.
\ref{Section:measures}.

\section{Two measures of quantum correlations}
\label{Section:measures}
As has been already mentioned, the two measures of quantum  correlations of our interest are the
 SR-entanglement (the traditional measure \cite{HW,Wootters})
  and the
 informational correlation  \cite{Z_inf} (the measure responsible for  eigenvector-parameters transfer).  
  Before proceed to the subject of this section we note the two features of the informational correlation: (i) it is discrete-valued and 
 (ii) its existence depends on the  set of   determinant conditions responsible for solvability of the 
 associated linear system of algebraic equations. These determinant conditions 
 are shown to be  appropriate objects to 
 compare with SR-entanglement.

\subsection{SR-entanglement as a traditional measure of quantum correlations}
\label{Section:C}
We consider the SR-entanglement  using the Wootters criterion 
\cite{HW,Wootters}
 taking the initial state of the sender and receiver in  form (\ref{inSR}-\ref{UR}) 
and the initial state of the transmission line $\rho^{TL}$ in  ground state  (\ref{TL}).

 Since  the entanglement $E$  is a monotonic function of so-called 
concurrence $C$, $E=-\frac{1+\sqrt{1-C^2}}{2}\log_2 \frac{1+\sqrt{1-C^2}}{2}
-\frac{1-\sqrt{1-C^2}}{2}\log_2 \frac{1-\sqrt{1-C^2}}{2}$,
 we base our 
consideration on this quantity, which can be calculated as follows:
\begin{eqnarray}\label{Conc}
C=\max(0, 2 \lambda_{max} - \sum_{i=1}^4 \lambda_i), \;\;
\lambda_{max} = \max(\lambda_1,\lambda_2,\lambda_3,\lambda_4),
\end{eqnarray}
where $\lambda_i$ are the eigenvalues of the following matrix
\begin{eqnarray}
\tilde\rho^{(SR)} = \sqrt{\rho^{SR} (\sigma_y \otimes \sigma_y)(\rho^{SR})^* 
(\sigma_y \otimes \sigma_y)}, \;\;\sigma_y=\left(
\begin{array}{cc}
0&-i\cr
i&0\end{array}\right).
\end{eqnarray}

\subsection{Informational correlation}
\label{Section:InfCor}
The informational correlation defined  in \cite{Z_inf} 
is the number of independent  eigenvector-parameters of the sender's initial density matrix  which can be 
registered at the receiver at some time instant $t$. We briefly recall its features. 
{The sender's  initial density matrix $\rho^S_0$ and the receiver's density matrix $\rho^R$ at some time instant $t$  can be written, respectively,    
 as follows:}
\begin{eqnarray}
&&
\rho^S_0=\left(
\begin{array}{cc}
1-x_1 & x_2 + i x_3\cr
x_2- i x_3 & x_1
\end{array}
\right)
,\\\nonumber
&&
\rho^R(t,x) = {{\mbox{Tr}}_{S,TL} \rho(t,x) }= \left(
\begin{array}{cc}
1-y_1(t,x) & y_2(t,x) + i y_3(t,x)\cr
y_2(t,x)- i y_3(t,x) & y_1(t,x)
\end{array}
\right),
\end{eqnarray}
{
where $x=(x_1,x_2,x_3)$ and  the  trace is taken over the nodes of sender $S$ 
and transmission line $TL$ (i.e., over all the nodes except for the $N$th one). }
Here, in view of formulas { (\ref{inSR}) - (\ref{US})},
\begin{eqnarray}\label{x}
&&
x_1=\frac{1}{2}\left(1+(1-2\lambda^S) \cos(\alpha_1\pi) \right),\;\;
x_2=-\frac{1}{2}(1-2\lambda^S) \sin(\alpha_1\pi) \cos(2 \alpha_2\pi) ,\;\;\\\nonumber
&&
x_3=\frac{1}{2}(1-2\lambda^S) \sin(\alpha_1\pi) \sin(2 \alpha_2\pi),
\end{eqnarray}
and $y_i$  depend explicitly on $x_i$, $i=1,2,3$ (see  Appendix, 
Sec.\ref{Section:appendixB}, for details):
\begin{eqnarray}\label{y}
y_1&=&\rho^{R}_{1;1} = T_{1 0;1 0} +(T_{1 1;1 1}-T_{1 0;1 0}) x_1 +2 Re(T_{1 0;1 1}) x_2 
- 2 Im (T_{1 0;1 1}) x_3 ,\\\nonumber
y_2&=&Re(\rho^{R}_{0;1}) = Re(T_{0 0;1 0}) +Re(T_{0 1;1 1}-T_{0 0;1 0}) x_1 +Re(T_{0 0;1 1} + T_{0 1;1 0}) x_2 
-\\\nonumber
&&Im (T_{0 0;1 1} - T_{0 1;1 0}) x_3,\\\nonumber
y_3&=&Im(\rho^{R}_{0;1}) = Im(T_{0 0;1 0}) +Im(T_{0 1;1 1}-T_{0 0;1 0}) x_1 +Im(T_{0 0;1 1} + T_{0 1;1 0}) x_2 
+\\\nonumber
&&Re (T_{0 0;1 1} - T_{0 1;1 0}) x_3,
\end{eqnarray}
where $T$-parameters are defined by the interaction   Hamiltonian, which is shown in Appendix, Secs.\ref{Section:appendixA} and  \ref{Section:appendixB}. { Remember, that the senders' and receiver's initial density matrices $\rho^S_0$, $\rho^R_0$  have
the forms  given in (\ref{inSR}) -- (\ref{UR}). Therefore the functions $y_i$ depend on $\alpha_j$ (through $x_i$, $i=1,2,3$) and 
  also on $\beta_j$, $j=1,2$, that will be used  in Sec.\ref{Section:lambda20}. }

\subsubsection{Determinant conditions quantifying   informational correlation}

In our case, there are two { eigenvector} control-parameters of the sender: $\alpha_i$, $i=1,2$. 
To  extract  the parameters $\alpha_i$ from the receiver's 
density matrix $\rho^R$, we have to  solve  system (\ref{y}), where $x_i$, $i=1,2,3$, are related
with  $\alpha_i$, $i=1,2$, by  formulas 
 (\ref{x}).
Obviously, the informational correlation $E^{SR}$ can take three  values:  0, 1, or 2.

 $\mathbf{E^{SR}=2.}$
In this case  system (\ref{y}) must be solvable for both parameters 
$\alpha_1$ and $\alpha_2$, so that the following  determinant condition must be satisfied:
\begin{eqnarray}\label{Delta2}
\Delta^{(2)}=\frac{1}{\Delta^{(2)}_0}\sum_{{n,m=1}\atop{m>n}}^3\sum_{{i,j=1}\atop{j>i}}^3
\left|\frac{\partial(y_i,y_j)}{\partial(x_n,x_m)}\right|
\left|
\frac{\partial(x_n ,x_m)}{\partial(\alpha_1,\alpha_2)}\right|\neq 0.
\end{eqnarray}

 $\mathbf{E^{SR}=1.}$
In this case system (\ref{y}) must be solvable for one 
of the parameters, either  $\alpha_1$ or $\alpha_2$, 
so that the following  determinant condition must be satisfied:
\begin{eqnarray}\label{Delta1}
\Delta^{(1)}=\frac{1}{\Delta^{(1)}_0}\sum_{i,n=1}^3\left|
\frac{\partial y_i}{\partial x_n}\right|
\left(\left|\frac{\partial x_n }{\partial \alpha_1} \right| +
\left|\frac{\partial x_n }{\partial \alpha_2} \right|\right) \neq 0.
\end{eqnarray}

$\mathbf{E^{SR}=0.}$ The both determinants $\Delta^{(i)}$, $i=1,2$, in (\ref{Delta2}) and 
(\ref{Delta1}) are identical to zero, so that no parameters 
can be registered at the receiver. 

In eqs.(\ref{Delta2}) and (\ref{Delta1}),
 the normalizations factors $\Delta^{(i)}_0$, $i=1,2$,  are defined by  the condition that 
$\langle\Delta^{(i)}\rangle_\Gamma=1$ if $y_j=x_j$, $j=1,2,3$ (in this case the
receiver's density matrix $\rho^R$ coincides 
with the sender's initial density matrix). Here  $\Gamma$ is the list of all the  parameters of the initial state,
\begin{eqnarray}\label{Gamma}
\Gamma=\{\lambda^S,\lambda^R,\alpha_1,\alpha_2,\beta_1,\beta_2\}.
\end{eqnarray}
Thus,
\begin{eqnarray}\label{normD}
&&
\Delta^{(2)}_0 =  \left\langle\sum_{{n,m=1}\atop{m>n}}^3
\left|
\frac{\partial(x_n, x_m)}{\partial(\alpha_1,\alpha_2)}\right|\right\rangle_{\lambda^S,\alpha_1,\alpha_2} =
\frac{\pi}{2},\\\nonumber
&&
\Delta^{(1)}_0 =
\left\langle\sum_{n=1}^3\left(\left|\frac{\partial x_n }{\partial \alpha_1} \right| +
\left|\frac{\partial x_n }{\partial \alpha_2} \right|\right) \right\rangle_{\lambda^S,\alpha_1,\alpha_2}=
\frac{1}{4}+\frac{3}{\pi},
\end{eqnarray}
In expressions (\ref{normD}),  we take into account that 
 all the  parameters $\Gamma$  have the same variation interval from 0 to 1 and the receiver's control 
 parameters $\lambda^R$, $\beta_i$, 
 $i=1,2$ do not appear in expressions (\ref{normD}), according to definitions of $x_i$ (\ref{x}).
 
Although the informational correlation $E^{(SR)}$  takes discrete values, 
it is directly related to the 
determinants $\Delta^{(i)}$ which, in turn, are the usual continuous
functions of the time and the control parameters. 
These determinants $\Delta^{(i)}$ are the functions which we deal with hereafter.  
Some useful properties of the determinants are given in Appendix, Sec.\ref{Section:appendixC}.

\section{A particular model of communication line based on spin-1/2 chain of $N=40$ nodes}
\label{Section:example}

We consider the  evolution of a homogeneous spin-1/2 chain of $N=40$ nodes governed 
by the nearest neighbor XY-Hamiltonian 
 \begin{eqnarray}\label{XY}
H=  \sum_{i=1}^{39} D(I_{ix} I_{(i+1)x} + I_{iy} I_{(i+1)y})
,
\end{eqnarray}
where  $D$ is   the coupling constant between the 
nearest neighbors, $I_{j\alpha}$ ($j=1,\dots,40$, $\alpha=x,y,z$) is the 
$j$th spin projection on the $\alpha$-axis.
In our model, we use the dimensionless time $Dt$  formally setting $D=1$. 
Studying the correlations among the sender and receiver, 
it is natural to consider the time  instant $t$ such that 
the  SR-concurrence $C$ and/or 
the determinants  $\Delta^{(i)}$  averaged over the 
initial conditions (i.e., $\langle \bar C\rangle_{\lambda^S,\lambda^R}$  and/or 
$\langle \bar  \Delta^{(i)}\rangle_{\lambda^S,\lambda^R}$) 
are maximal.
For $N=40$, this time instant   $t=43.442$ is found in 
Appendix, Sec.\ref{Section:appendix0}.

\subsection{SR-concurrence as a function of  
control parameters}
\label{Section:lambda0}
In this section we represent the detailed analysis of the { SR-concurrence}
as a function of control parameters. 
 First of all, we  calculate  the SR-concurrence averaged over 
 all the eigenvector parameters $\tilde\Gamma$  and represent such mean  SR-concurrence   as a function of 
 the eigenvalues $\lambda^S$ and $\lambda^R$ in Sec.\ref{Section:lambda}. 
After that, the effects of the control parameters $\tilde\Gamma$ { on the SR-concurrence} will be demonstrated 
in terms of the standard deviations with respect to these parameters in Sec.\ref{Section:angles}. 
Additional details are discussed in Secs.\ref{Section:concFeat}-\ref{Section:nn}.

\subsubsection{Mean SR-concurrence $\bar C$  in dependence on initial eigenvalues}
\label{Section:lambda}
\begin{figure*}
   \epsfig{file=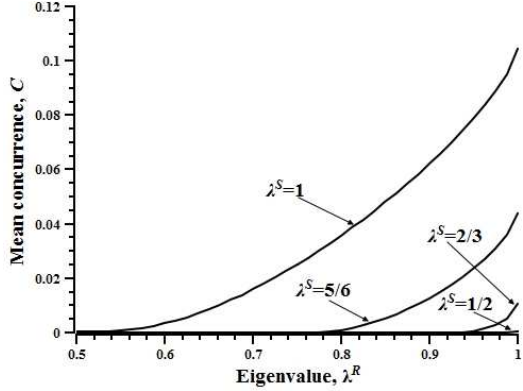,
  scale=0.4
   ,angle=0
}
\caption{ 
The mean SR-concurrence $\bar{C}$
as a function of $\lambda^S$ 
and $\lambda^R$. 
Each  line corresponds to the particular 
value of $\lambda^S = \frac{1}{2},\frac{2}{3},\frac{5}{6},1$ 
increasing from the bottom to the top of figure. 
 The gridding for averaging: 
 $\alpha_i , \beta_i= 0.05 n$, $n=0,1,\dots,20$.
The same gridding is used in Fig.\ref{Fig:disp1C}. 
We show only the region $\frac{1}{2}\le\lambda^S\le1$, $\frac{1}{2}\le\lambda^R\le1$ because of the symmetry $\lambda\leftrightarrow 1-\lambda$.
} 
  \label{Fig:avrC} 
\end{figure*}

To calculate the mean SR-concurrence  $\bar{C}$    averaged over the parameters $\tilde \Gamma$ 
we  use formula (\ref{avF}) with the substitution 
$F=C$. Therewith,  
$C$  is  defined  in eq.(\ref{Conc}).
The mean SR-concurrence  $\bar{C}$  as a function of $\lambda^S$ and $\lambda^R$ is
 depicted in Fig. \ref{Fig:avrC}. In this figure, each 
 curve corresponds to a particular value of $\lambda^S=\frac{1}{2}, \frac{2}{3},\frac{5}{6},1$,
 while $\lambda^R$ is along the abscissa axis (remember the symmetry $S\leftrightarrow R$). 
First of all we shell note that 
the mean SR-concurrence decreases with an increase in the chain length. It is also an increasing function of both $\lambda^R$ and $\lambda^S$ reaching its maximal value $\bar{C}_{max}$ at 
$\lambda^S=\lambda^R=1$. For the chain of $N=40$ nodes, this maximum is $\bar{C}_{max}=1.15\times 10^{-1}$. Next, the  value of $\bar{C}$ tends to zero 
as $\lambda_S \to \frac{1}{2}$ 
(or $\lambda_R \to \frac{1}{2}$) with $\bar{C}|_{\lambda_S=\frac{1}{2},\lambda_R=1} = 
\bar{C}|_{\lambda_S=1,\lambda_R=\frac{1}{2}}=1.87\times 10^{-4}$. 

We also observe that all the curves (except  the curve $\lambda^S=1$) start from some $\lambda^R>\frac{1}{2}$. This means that there is a 
region on the plane of the control parameters $(\lambda^R,\lambda^S)$, which maps into the states of the subsystem $SR$ with zero SR-entanglement regardless of the 
eigenvector control parameters $\tilde\Gamma$. This property of our model will be discussed in Sec.\ref{Section:concFeat} in more details.

\subsubsection{Effect of eigenvector initial parameters $\tilde \Gamma$}
\label{Section:angles}

\begin{figure*}
   \epsfig{file=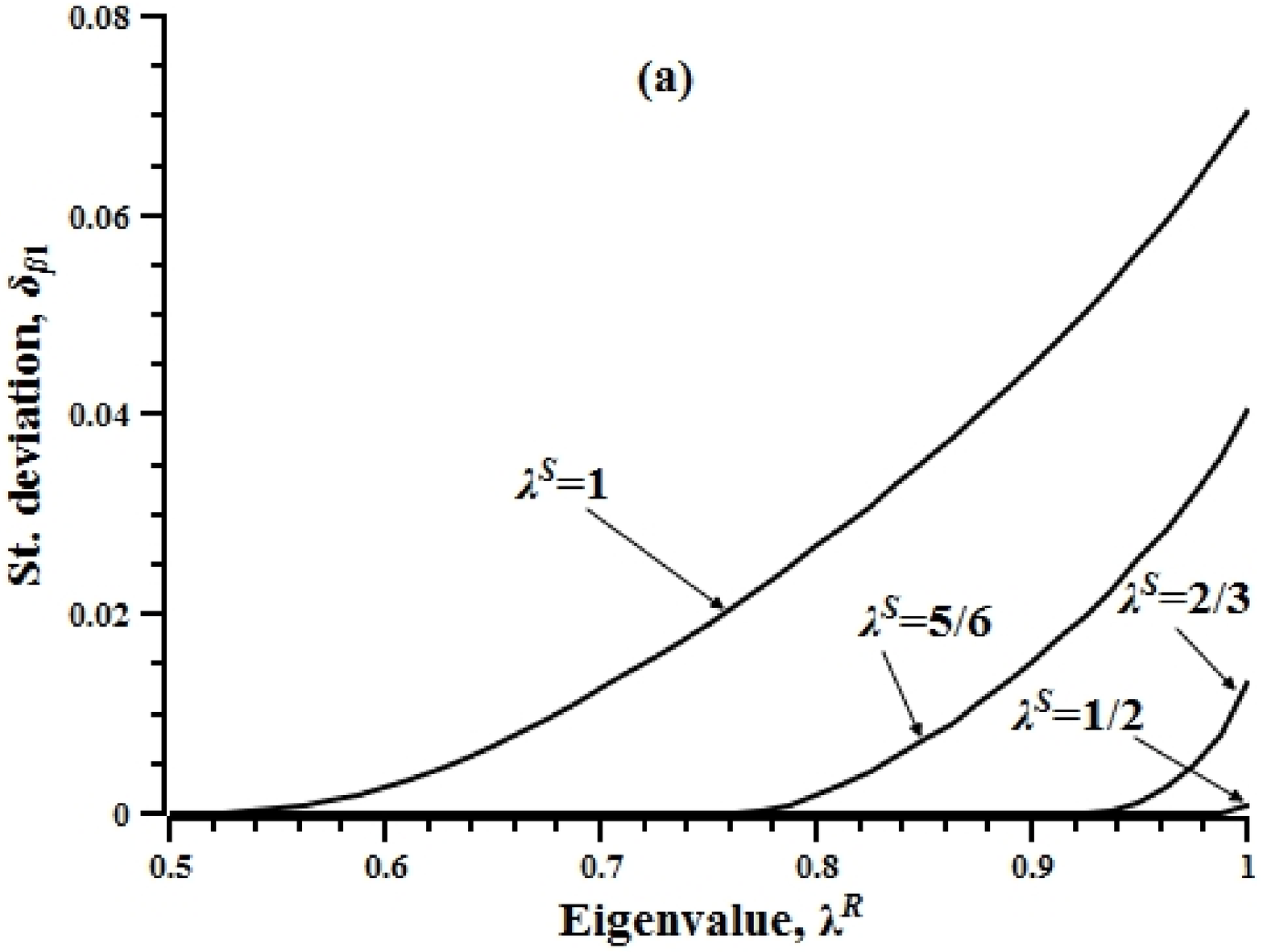,
  scale=0.45
   ,angle=0
}
\epsfig{file=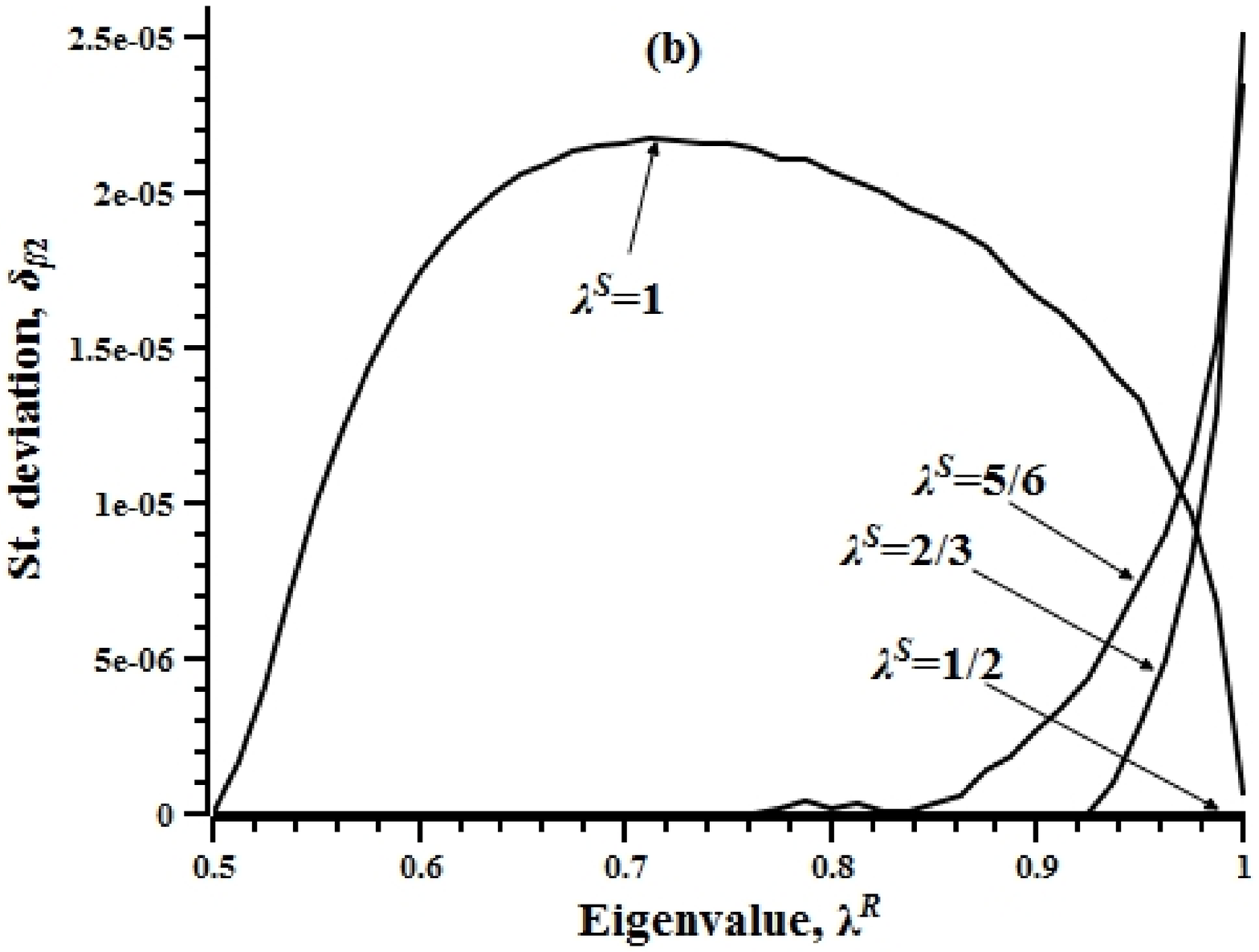,
  scale=0.45
   ,angle=0
}
\caption{ 
 The standard deviations $\delta^{(C)}_{\beta_i}$, $i=1,2$,  as
functions of $\lambda^S$ and $\lambda^R$. Each  line corresponds to the particular 
value of $\lambda^S = \frac{1}{2}, \frac{2}{3},\frac{5}{6},1$ 
increasing from the bottom to the top of figure.  Both standard deviations vanish
at $\lambda^R=\frac{1}{2}$.
(a) $\delta^{(C)}_{\beta_1}$ with the  maximal value
$\delta^{(C)}_{\beta_1}|_{\lambda^S=\lambda^R=1} = 7.05\times 10^{-2}$.
(b)  $\delta^{(C)}_{\beta_2}$ with the maximal value 
$\delta^{(C)}_{\beta_2}|_{{\lambda^R=0.75}\atop{\lambda^S=1}} =2.16\times 10^{-5} $, 
}
  \label{Fig:disp1C} 
\end{figure*}

The effect of $\alpha_i$  on $\bar{C}$ is quite  similar to that of $\beta_i$, $i=1,2$, owing 
to the symmetry $S\leftrightarrow R$  of the system. Therefore,  hereafter in this section we consider only two 
standard deviations of the SR-concurrence with respect to the parameters $\beta_i$, $i=1,2$, using   formula (\ref{delta}) with the substitutions $F=C$ and 
$\gamma = \beta_i$:
\begin{eqnarray}\label{deltaC}
\delta^{(C)}_{\beta_i} = \sqrt{\Big\langle \Big(\bar{C} - 
\langle C \rangle_{\tilde\Gamma_{\beta_i}}\Big)^2 \big\rangle_{\beta_i}},
\;\;i=1,2.
\end{eqnarray}
These standard deviations  
are shown
in Fig.\ref{Fig:disp1C}. 
Fig.\ref{Fig:disp1C}a and Fig.\ref{Fig:disp1C}b
demonstrate, respectively,  that 
$\bar C$ is sensitive to the parameter $\beta_1$, 
and $\bar C$  is  not sensitive to the parameter $\beta_2$. For this reason,   the parameters $\alpha_1$, $\beta_1$
and $\alpha_2$, $\beta_2$ are referred to as, respectively, strong and weak parameters. We can 
  neglect the effect of $\beta_2$ and $\alpha_2$  putting $\beta_2=\alpha_2=0$ in most calculations. 

The $\lambda$-dependence of standard deviations $\delta^{(C)}_{\beta_1}$ is quite similar to that of the mean concurrence. 
The function  $\delta^{(C)}_{\beta_2}$
is different and it is not monotonic  with respect to $\lambda^R$ and $\lambda^S$, which is shown in Fig. \ref{Fig:disp1C}b.
 Therewith, unlike the mean concurrence, 
both $\delta^{(C)}_{\beta_1}$ and $\delta^{(C)}_{\beta_2}$ vanish  at $\lambda^R=\frac{1}{2}$, 
because in this case the receiver's initial density matrix is proportional to the identity matrix and therefore does not depend on 
the  parameters $\beta_i$. 

Similar to the mean concurrence, each of the curves in Fig.\ref{Fig:disp1C}a and Fig.\ref{Fig:disp1C}b 
  starts from some
$\lambda^R>\frac{1}{2}$ (while the  curve $\lambda^S=1$ starts from $\lambda^R = \frac{1}{2}$), 
which means that there is a region on the $(\lambda^R,\lambda^S)$-plane corresponding to the
states of the subsystem $SR$ with  zero $\delta^{(C)}_{\beta_1}$ and $\delta^{(C)}_{\beta_2}$. Of cause,  the standard 
deviations are zero in the domain of   the $(\lambda^R,\lambda^S)$-plane where 
$\bar{C}=0$,  see  
Sec.\ref{Section:lambda}.
The pre-image of non-entangled states  in the control parameter space  together with its boundary  deserves 
the special consideration which is given in the next subsection.

\subsubsection{Pre-image of non-entangled  states $\rho^{SR} $  
in  control-parameter space and its boundary}
\label{Section:concFeat}
In this subsection we disregard the effect of weak control parameters setting 
 $\alpha_2=\beta_2=0$, which 
significantly simplifies the numerical calculations.

\begin{figure*}
   \epsfig{file=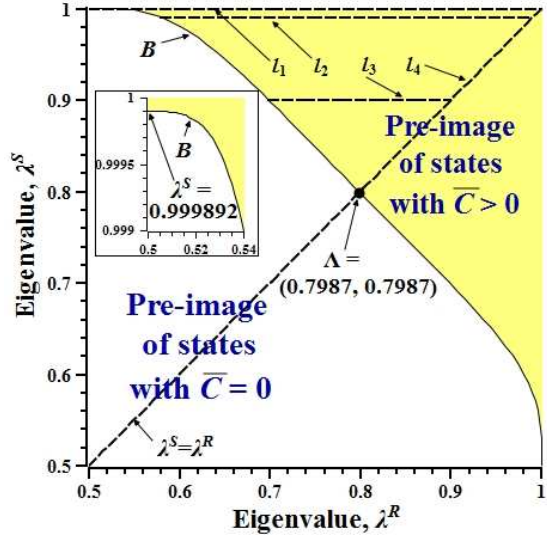,
  scale=0.45
   ,angle=0
}
\caption{ 
 The pre-images  of states  $\rho^{SR}$  with $\bar{C}=0$ and $\bar{C}>0$ separated by the boundary $B$ 
  on the plane $(\lambda^R,\lambda^S)$ at $t=43.442$. The inset represents the  boundary $B$ in the close neighborhood of 
  $\lambda^R=\frac{1}{2}$ showing  the  limiting value $\lambda^S_{min}=0.999892$ such that,
  if $\lambda^S>\lambda^S_{min}$, then $\bar{C}> 0$ for all $\lambda^R$, $0\le \lambda^R\le 1$.
  The boundary $B$ is symmetrical with respect to the bisectrix $\lambda^S=\lambda^R$, which 
  crosses the boundary at the point $\Lambda= (0.7987,0.7987)$.  The pairs of parameters $(\lambda^R,\lambda^S)$ along the 
  dashed lines $l_1$ -- $l_4$ (including the bisectrix) will be used for constructing the pre-images of entangled states on the $(\beta_1,\alpha_1)$-plane in Fig.\ref{Fig:W2}.   
}
 \label{Fig:zero} 
\end{figure*}

According to Figs.\ref{Fig:avrC} and  \ref{Fig:disp1C}, the  
SR-concurrence $C$ strongly depends on the control parameters $\lambda^S$, $\lambda^R$, $\alpha_1$ and $\beta_1$ and vanishes inside
of a large domain of these parameters. 
In particular, its mean value $\bar{C}$ vanishes  if the initial eigenvalues $\lambda^S$ and $\lambda^R$
 are inside of certain domain  on the plane 
 $(\lambda^R,\lambda^S)$.
 This domain  at $t=43.442$ is shown in Fig.\ref{Fig:zero} and is called the pre-image of non-entangled 
  states $\rho^{SR}$ ($C\equiv 0$) on the plane $(\lambda^R,\lambda^S)$.
 
We see that there is a well defined boundary (the line $B$) separating the pre-images   of the
states with $\bar{C}=0$ and $\bar{C}>0$. 
Furthermore, the inset  in this figure shows that there is  the
limiting value  $\lambda^S_{min}=0.999892$, such that if $\lambda>\lambda^S_{min}$, then the 
 mean SR-concurrence $\bar{C}> 0$ for all $\lambda^R$, $0\le\lambda^R\le1$. 

\begin{figure*}
   \epsfig{file=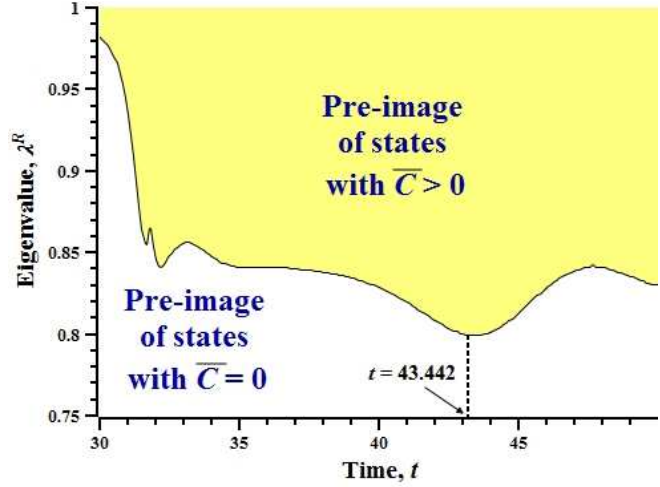,
  scale=0.5
   ,angle=0
}
\caption{ 
 The evolution of the point $\Lambda$ 
 of the boundary $B$  (see Fig.\ref{Fig:zero}) along the bisectrix $\lambda^S=\lambda^R$. 
 The position of $\Lambda$ on the bisectrix is defined by the  parameter $\lambda^R$ 
(ordinate axis).  
}
 \label{Fig:zeroEV} 
\end{figure*}

Obviously, the boundary $B$ on the $(\lambda^R,\lambda^S)$-plane  evolves in time keeping the symmetry with respect to the line 
$\lambda^S=\lambda^R$ which is  shown in Fig.\ref{Fig:zero}. 
To demonstrate this  evolution  we take a boundary point 
$\Lambda$   with the coordinates   $\lambda^S=\lambda^R=0.7987$ (marked in Fig.\ref{Fig:zero}) and show its evolution  along the bisectrix 
$\lambda^S=\lambda^R$  in Fig.\ref{Fig:zeroEV}. We see that
this point 
reaches its minimal position
at $t=43.442$
(the pre-image of entangled states is above the evolution curve), i.e., exactly at the time instant found  for state registration. 
Therefore, if the initial eigenvalues are taken inside of the domain below the boundary $B$ in Fig.\ref{Fig:zero}, 
then evolution can not create the entangled 
states irrespective of  the values of the eigenvector control-parameters $\alpha_i$ and $\beta_i$. 

\subsubsection{Witness of SR-entanglement}
\label{Section:withess}
\begin{figure*}
   \epsfig{file=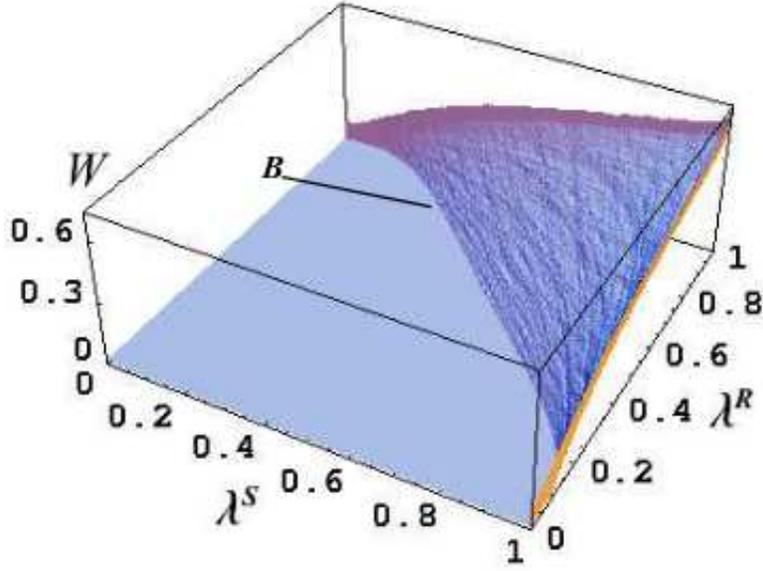,
  scale=1
   ,angle=0
}
\caption{ 
The  witness of SR-entanglement $W$ as a function of  $\lambda^R$ and $\lambda^S$ at $t=43.442$. 
}
 \label{Fig:W} 
\end{figure*}

The numerical simulations show that even if the initial eigenvalues are inside of the pre-image of 
 states with $\bar{C}>0$,
the SR-concurrence  equals  zero  in large domain of the control parameters 
 $\alpha_i$ and $\beta_i$, $i=1,2$. For clarity,  we introduce the following witness of SR-entanglement
\begin{eqnarray}
W(\lambda^S,\lambda^R) =\int \theta(\lambda^S,\lambda^R,\alpha_1,\beta_1) 
d \alpha_1 d\beta_1, 
\end{eqnarray}
where 
\begin{eqnarray}
\theta(\Gamma) =\left\{\begin{array}{ll}
1,& C(\lambda^S,\lambda^R,\alpha_1,\beta_1)>0\cr
0,& C(\lambda^S,\lambda^R,\alpha_1,\beta_1)=0
\end{array}\right.,
\end{eqnarray}
and  we take into account that all parameters $\tilde \Gamma$ vary inside of the unit interval (\ref{intervals}), see Fig.\ref{Fig:W}.

If the RS-concurrence were nonzero for all  values of $\alpha_1$ and $\beta_1$, then 
the graph would be horizontal plane with $z$-coordinate equal 1. However, our surface is different and 
it is always below the above mentioned plane. This means that
the SR-concurrence is zero inside of  large domain on the plane $(\alpha_1,\beta_1)$ and 
this domain depends on $\lambda^S$ and $\lambda^R$.
Moreover,  the slope of the surface in  Fig.\ref{Fig:W} shows that the area of  the pre-image of the entangled states 
on  the plane of  the parameters $\alpha_1,\beta_1$  increases with the distance 
from the boundary curve. 

\subsubsection{Pre-image of entangled states on $(\beta_1,\alpha_1)$-plane}
\label{Section:alpbet}
\begin{figure*}
   \epsfig{file=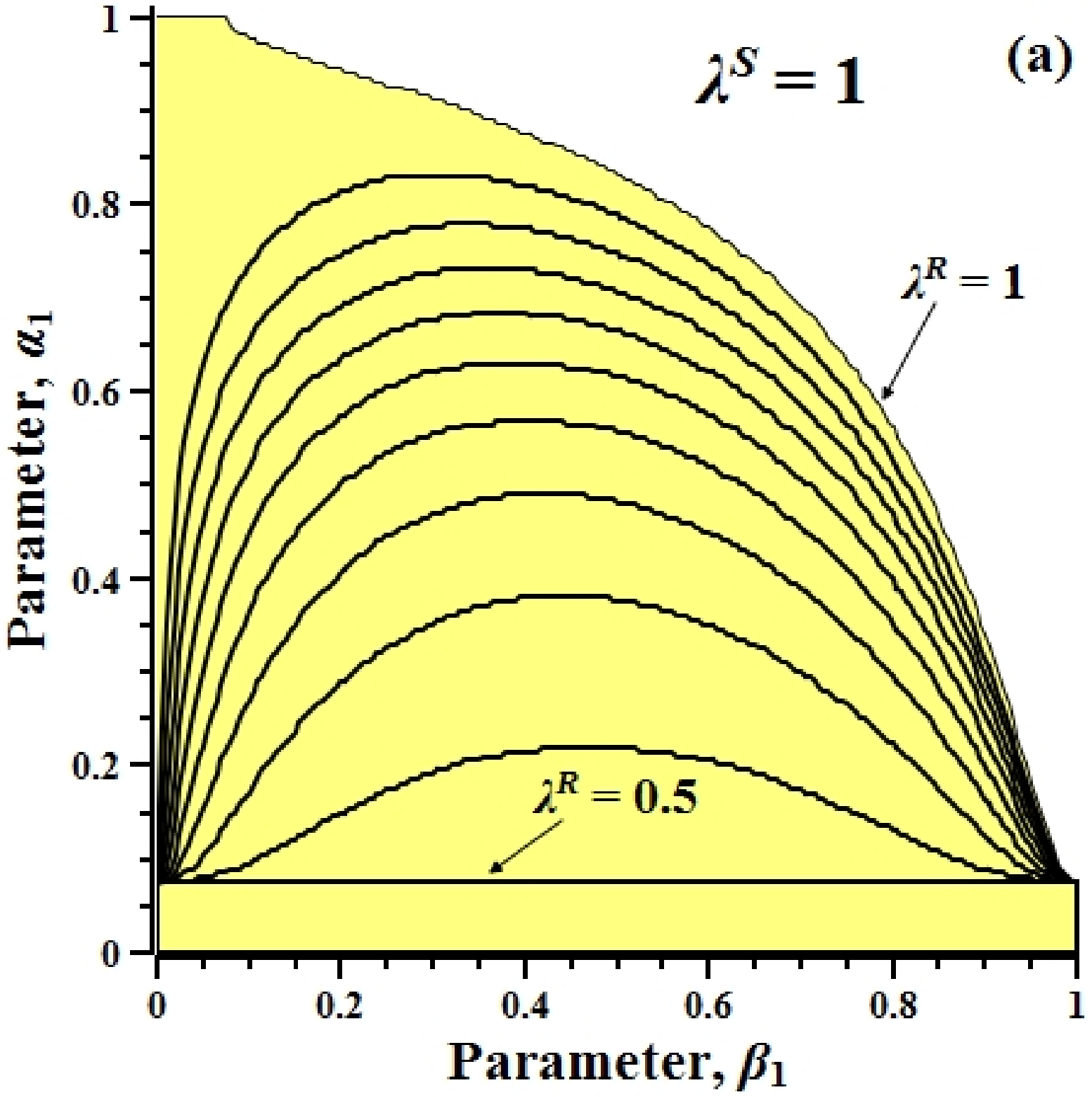,
  scale=0.4
   ,angle=0
}\epsfig{file=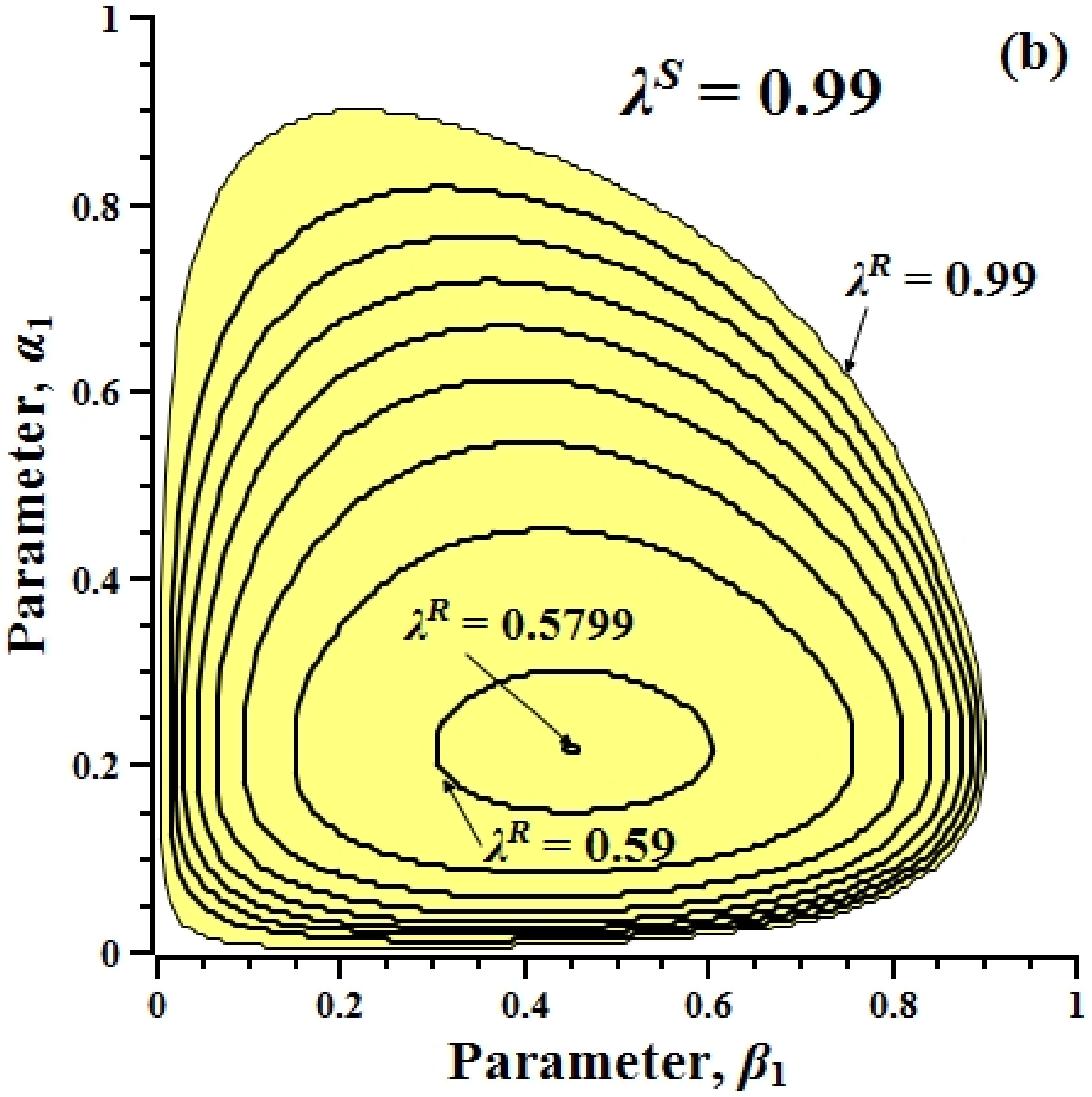,
  scale=0.4
   ,angle=0
}\\
\epsfig{file=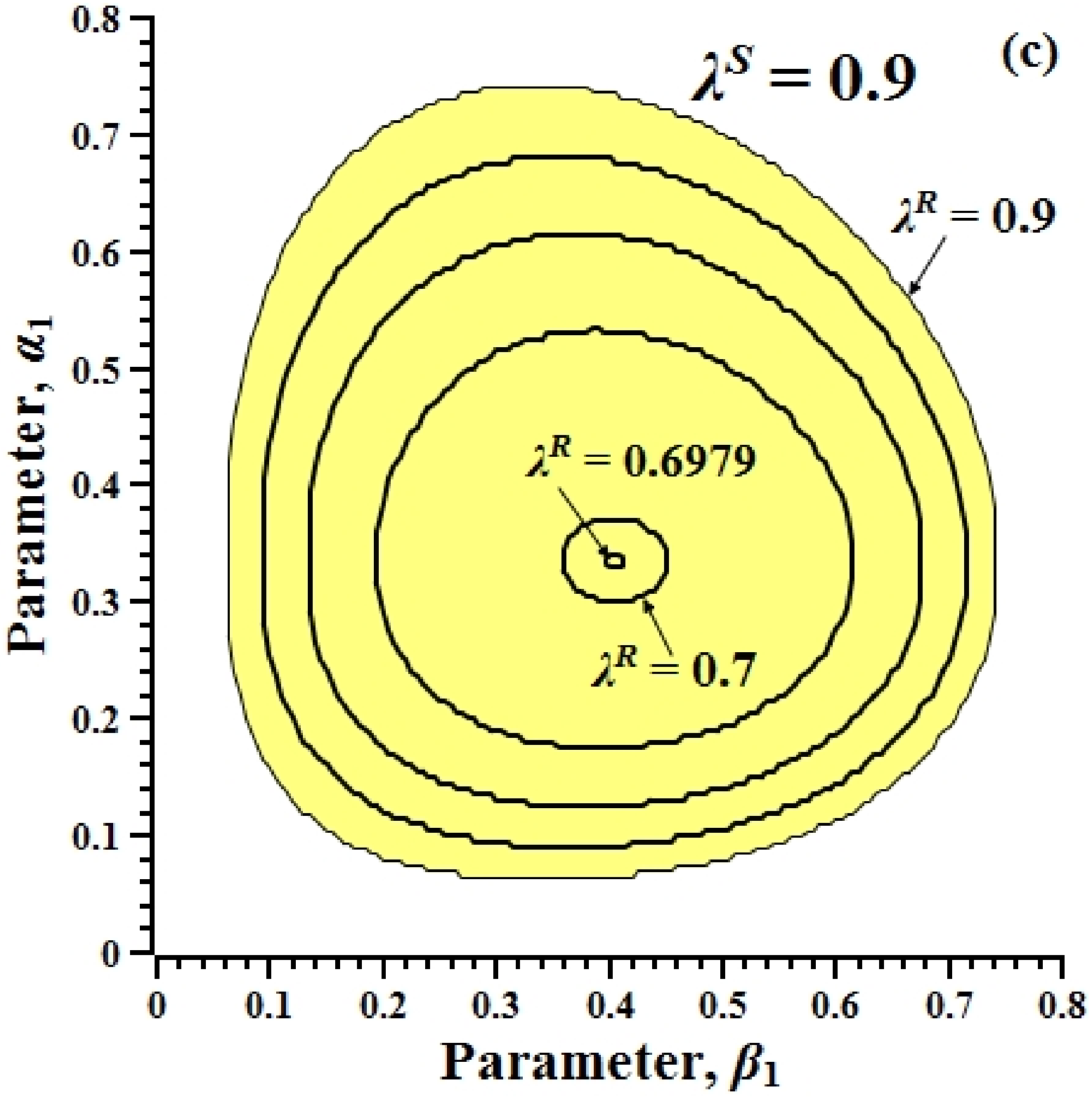,
  scale=0.4
   ,angle=0
}\epsfig{file=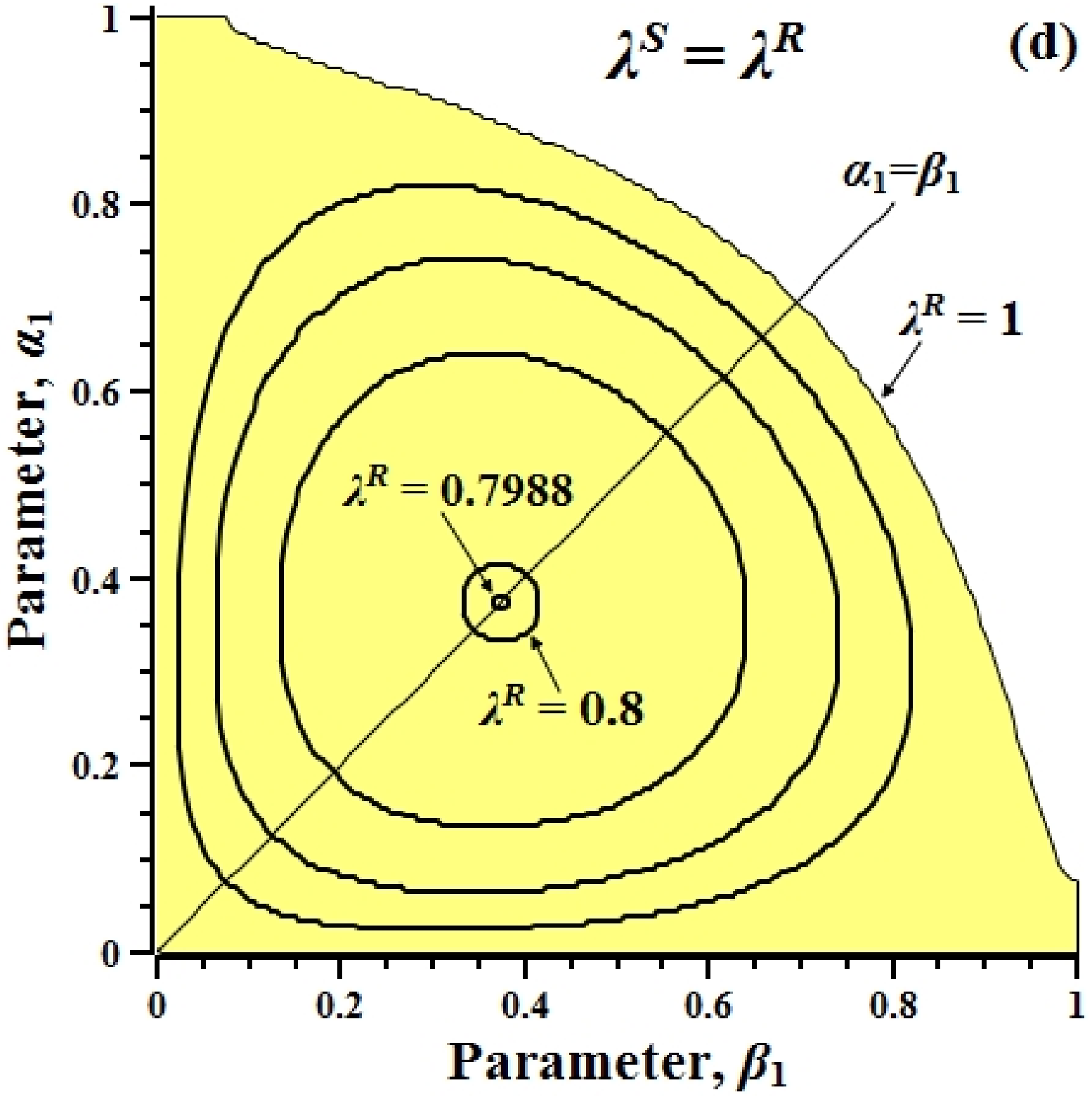,
  scale=0.4
   ,angle=0
}
\caption{ The pre-image of entangled states on the plane $(\beta_1,\alpha_1)$ for the eigenvalues $(\lambda^R,\lambda^S)$ 
 along the lines
$l_1$, $l_2$, $l_3$, $l_4$ (see Fig.\ref{Fig:zero}).
Each curve represents the boundary of the pre-image of entangled states associated with the  particular value of $\lambda^R$ on the appropriate line $l_k$; the $\lambda^R$-interval between the neighboring curves is 
$\Delta\lambda^R=0.05$ except  the central smallest curves which are specified.  All  pre-images of the entangled states are simply-connected domains. 
 (a) $\lambda^S=1$, $\frac{1}{2}\le \lambda^R \le1$  (the  line $l_1$ in Fig.\ref{Fig:zero}). 
 The lower rectangular curve  corresponds to $\lambda^R=\frac{1}{2}$, 
 the largest curve corresponds to  $\lambda^R=1$. 
 (b) $\lambda^S=0.99$,  $ 0.5799\le \lambda^R \le 0.99$ 
 (the line $l_2$ in Fig.\ref{Fig:zero}), the central curve  corresponds to $\lambda^R=0.5799$, 
 the next one corresponds to 
 $\lambda^R=0.59$, the largest curve corresponds to $\lambda^R= 0.99$.
 (c) $\lambda^S=0.9$, $ 0.6979\le \lambda^R \le 0.9$  (the  line $l_3$ in Fig.\ref{Fig:zero}),
 the central curve corresponds to $\lambda^R=0.6979$, 
 the next curve corresponds to $\lambda^R=0.7$, the largest   curve corresponds to $\lambda^R=0.9$,
(d) $\lambda^R= \lambda^S$,  $0.7988\le \lambda^R\le 1 $ (the line $l_4$ in Fig.\ref{Fig:zero}),  
the central  curve corresponds to $\lambda^R=0.7988$, 
the next one corresponds to $\lambda^R=0.8$, the  largest curve corresponds to $\lambda^R=1$. 
}
 \label{Fig:W2} 
\end{figure*}

From Fig.\ref{Fig:W} it follows that the pre-image of entangled states on the plane $(\beta_1,\alpha_1)$ 
depends on the particular values of $\lambda^S$ and $\lambda^R$. To demonstrate this dependence,
we consider the pre-images of the  entangled states on the plane $(\beta_1,\alpha_1)$, 
corresponding to different pairs  $(\lambda^R,\lambda^S)$  taken
 along the four lines  $l_1\;- \; l_4$ in Fig.\ref{Fig:zero} with 
 \begin{eqnarray}
 \lambda^S|_{l_1}=1,\;\;\lambda^S|_{l_2}=0.99,\;\;\lambda^S|_{l_3}=0.9,\;\;\lambda^S|_{l_4}=\lambda^R.
 \end{eqnarray}
The results are depicted in Fig.\ref{Fig:W2}, where  
each closed contour is the boundary of the  pre-image of entangled states associated with a particular 
value of the pair $(\lambda^R,\lambda^S)$. 
The smallest central  contour (which almost shrinks to a  point) in each of Figs. \ref{Fig:W2}b-d corresponds to the point on the appropriate line
approaching  the boundary $B$ from the right. 

The line $l_1$ (see Fig.\ref{Fig:W2}a)   corresponds to the 
maximal eigenvalue $\lambda^S=1$, so that  $\lambda^R$ runs all the values, $0\le \lambda^R\le 1$. 
This figure shows us that there is a rectangular subregion 
of the pre-image of entangled states for any $\lambda^R$ ($0\le \lambda^R\le 1$): 
$ 0\le \beta_1 \le 1,\;\; 0\le \alpha_1\le 0.0763$.
If, in addition, $\lambda^R=1$, then there are two rectangular subregions 
of the pre-image of the entangled states:
\begin{eqnarray}\label{rectangle}
&&{\mbox{1st subregion}}:\;\; 0\le \beta_1 \le 1,\;\; 0\le \alpha_1\le 0.0763,\\\nonumber
&&{\mbox{2nd subregion}}:\;\; 0\le \beta_1 \le 0.0763,\;\; 0\le \alpha_1<1.
\end{eqnarray}
The 1st rectangular subregion  is the pre-image of states with $\bar{C}>0$ corresponding  to $\lambda^R=\frac{1}{2}$, i.e., the initial state of the receiver
is proportional to the identity matrix and, consequently, doesn't depend on the parameters 
$\beta_i$. 
In this case, the SR-entanglement disappears if $\alpha_1>0.0763$.  Similarly, the 2nd rectangular subregion is the pre-image of states with $\bar{C}>0$ corresponding   to $\lambda^S=\frac{1}{2}$, 
the SR-entanglement disappears if $\beta_1>0.0763$.
Figs.\ref{Fig:W2}b-d  clearly demonstrate that the  area of the pre-image of the entangled  
states increases with the distance from the boundary $B$ which agrees with Fig.\ref{Fig:W}. 
Remark that the line, corresponding to $\lambda^R=\lambda^S=1$, appears in 
two figures: Fig.\ref{Fig:W2}a and Fig.\ref{Fig:W2}d.

\subsubsection{Neighborhood of boundary $B$}
\label{Section:nn}
\begin{figure*}
   \epsfig{file=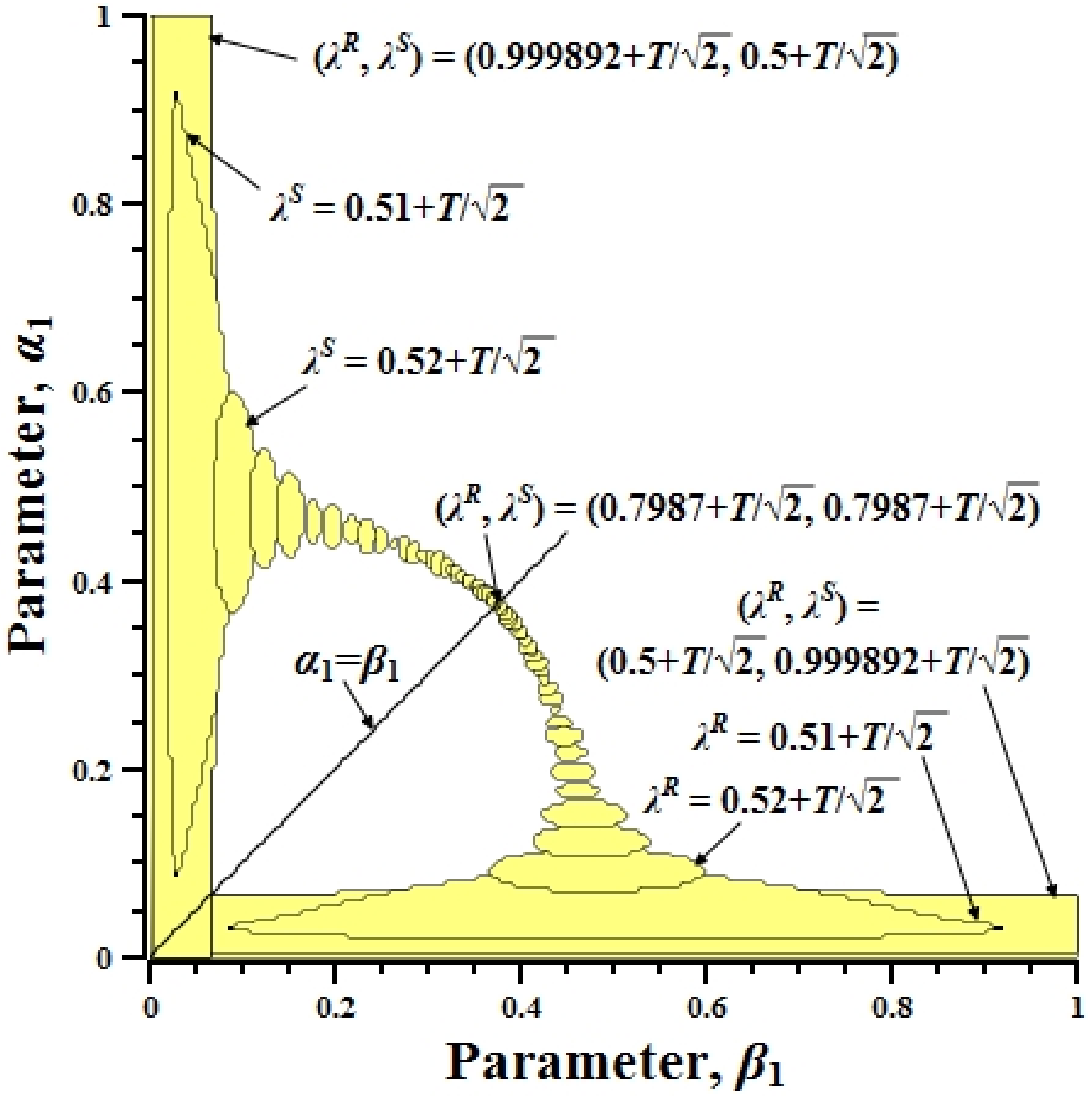,
  scale=0.5
   ,angle=0
}
\caption{ The  formation of SR-entanglement on the line neighboring the boundary $B$ in Fig.\ref{Fig:zero}. 
This neighboring line is obtained by  shifting the boundary $B$  
along the bisectrix  $\lambda^S=\lambda^R$ over $T=0.0001$. Thus, the rectangular contours 
above the abscissa axis and near the ordinate axis  correspond, respectively,  to 
the points $(\lambda^R,\lambda^S) = (0.5+\frac{T}{\sqrt{2}},0.999892+\frac{T}{\sqrt{2}} )$ and 
$(\lambda^R,\lambda^S) = (0.999892+\frac{T}{\sqrt{2}},0.5+\frac{T}{\sqrt{2}} )$. The crosspoint with the bisectrix is
$(\lambda^R,\lambda^S) =(0.7987+\frac{T}{\sqrt{2}},0.7987+\frac{T}{\sqrt{2}})$. For the  two neighboring 
curves $i$th and $(i+1)$th above and below  the bisectrix  $\alpha_1=\beta_1$, we have, respectively, 
$|\lambda^S_{i+1} -\lambda^S_{i}| =0.01$,
and $|\lambda^R_{i+1} -\lambda^R_{i}| =0.01$. 
}
 \label{Fig:Bet} 
\end{figure*}

 Finally, we represent the nearest neighborhood of the  boundary $B$ on the plane $(\alpha_1,\beta_1)$ in Fig.\ref{Fig:Bet}. 
 This neighborhood corresponds to the line which defers from the boundary line $B$ by the   shift $T=0.0001$ along the positive direction of the bisectrix
 $\lambda^S=\lambda^R$. This shifted line can be considered as a line  of formation of entanglement.
 
 \subsection{Informational correlation as  function of  
control parameters}
 \label{Section:lambda20}
 
 Analyzing the results of Sec.\ref{Section:lambda0} we conclude that the SR-concurrence is ''selective'' to the values of control parameters and vanishes in large 
 domain of their space.  In this section, we show that the informational correlation behaves quite differently.

 We study the  informational correlation $E^{SR}$ following the strategy of Sec.\ref{Section:C} and base our 
consideration on the determinants $\Delta^{(i)}$ instead of $E^{SR}$ itself because they  are   responsible
for the  registration of the sender's control parameters $\alpha_i$, $i=1,2$, at the receiver. 
First, we consider  the mean  determinants as the  functions of eigenvalues $\lambda^S$ and $\lambda^R$,
Sec.\ref{Section:lambda2}, and then we turn to the effects of the  eigenvector control parameters in Sec.\ref{Section:eigenvectors2}.

\subsubsection{Mean determinants   in dependence on initial eigenvalues}
\label{Section:lambda2}
To calculate the mean  determinants  $\bar{\Delta}^{(i)}$ we use formula (\ref{avF}) with substitutions 
$F=\Delta^{(i)}$, $i=1,2$. Therewith,  
  $\Delta^{(2)}$ and  $\Delta^{(1)}$ are  defined, 
respectively,  in eqs. (\ref{Delta2}) and 
(\ref{Delta1}). 

\begin{figure*}
   \epsfig{file=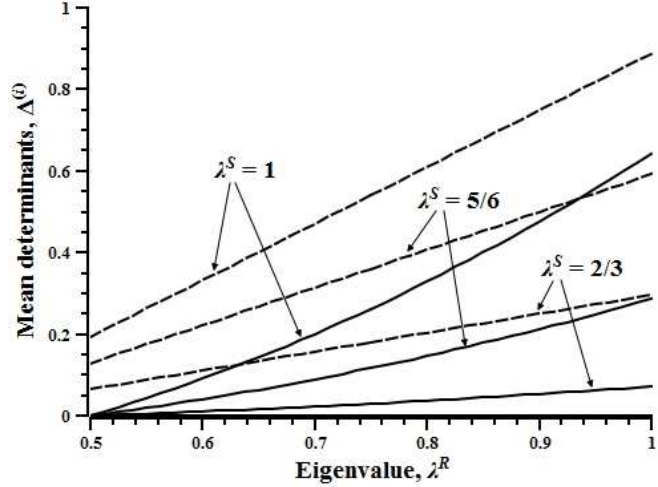,
  scale=0.5
   ,angle=0
}
\caption{ 
The mean determinants  $\bar{\Delta}^{(2)}$ (solid line)
and  $\bar{\Delta}^{(1)}$ (dashed line)
as functions of $\lambda^S$ 
and $\lambda^R$. 
Each  line corresponds to the  particular 
value of $\lambda^S = \frac{1}{2}, \frac{2}{3},\frac{5}{6},1$ 
increasing from the bottom to the top of the figure with   
$\bar{\Delta}^{2}|_{\lambda^S=\frac{1}{2}}=
\bar{\Delta}^{1}|_{\lambda^S=\frac{1}{2}}=0$. 
Each  function  $\bar{\Delta}^{(2)}$ and $\bar{\Delta}^{(1)}$
takes its  maximal value  at the boundary point  $\lambda^S=\lambda^R=1$: 
$\bar{\Delta}^{(2)}_{\max}=0.6413$ and  $\bar{\Delta}^{(1)}_{\max}=0.8869$. Therewith, $\bar{\Delta}^{(2)}$ turns to zero at   
$\lambda^R=\frac{1}{2}$  for all $\lambda^S$.
On the contrary, $\bar{\Delta}^{(1)}$  doesn't vanish at $\lambda^R=\frac{1}{2}$, its maximal value in this case is at $\lambda^S=1$: 
$ \bar{\Delta}^{(1)}|_{{\lambda^S=1}\atop{\lambda^R=\frac{1}{2}}}  = 0.1929$.
} 
  \label{Fig:avr} 
\end{figure*}

Calculating $\bar{\Delta}^{(i)}$, we have to take into account that the dependence of determinants 
 on $\alpha^{(i)}$, $\lambda^S$ is separated from their dependence on 
$\beta^{(i)}$, $\lambda^R$ in formulas (\ref{Delta2}) and 
(\ref{Delta1}) (see  Appendix, Sec.\ref{Section:propdet} for details).  Consequently,  formula (\ref{avF}) for the mean determinants $\Delta^{(i)}$ yields:
\begin{eqnarray}\label{Delta22}
\bar{\Delta}^{(2)}(\lambda^S,\lambda^R)=
\frac{1}{\Delta^{(2)}_0}\sum_{{n,m=1}\atop{m>n}}^3\sum_{{i,j=1}\atop{j>i}}^3
\left\langle\left|\frac{\partial(y_i,y_j)}{\partial(x_n,x_m)}
\right|\right\rangle_{\beta_1,\beta_2}
\left\langle\left|
\frac{\partial(x_n ,x_m)}{\partial(\alpha_1,\alpha_2)}\right|
\right\rangle_{\alpha_1,\alpha_2},
\end{eqnarray}
and
\begin{eqnarray}\label{Delta12}
\bar{\Delta}^{(1)}(\lambda^S,\lambda^R)=\frac{1}{\Delta^{(1)}_0}\sum_{i,n=1}^3\left\langle\left|
\frac{\partial y_i}{\partial x_n}\right|\right\rangle_{\beta_1,\beta_2}
\left\langle\left(\left|\frac{\partial x_n }{\partial \alpha_1} \right| +
\left|\frac{\partial x_n }{\partial \alpha_2} \right|\right)\right\rangle_{\alpha_1,\alpha_2} 
\end{eqnarray}
Moreover, the averaging over $\alpha_i$ can be simply done analytically, i.e.,
\begin{eqnarray}
\left\langle \left|
\frac{\partial(x_n, x_m)}{\partial(\alpha_1,\alpha_2)}
\right|\right\rangle_{\alpha_1,\alpha_2}
=  \frac{\pi}{2}(1-2\lambda^S)^2,\;\;(n,m)=(1,2), (1,3), (2,3)
\end{eqnarray}
and 
\begin{eqnarray}
\left\langle\left(\left|\frac{\partial x_n }{\partial \alpha_1} \right| +
\left|\frac{\partial x_n }{\partial \alpha_2} \right|\right)\right\rangle_{\alpha_1,\alpha_2}= \left\{
\begin{array}{ll}
|1-2\lambda^S|,& n=1\cr
\frac{6}{\pi}|1-2\lambda^S|, & n=2,3.
\end{array}
\right.
\end{eqnarray}
The mean determinants  $\bar{\Delta}^{(i)}$, $i=1,2$,  as  functions of $\lambda^S$ and $\lambda^R$ are
 depicted in Fig. \ref{Fig:avr}

We see  that, 
unlike the concurrence,  there is no domain on the plane $(\lambda^R,\lambda^S)$ resulting in the zero mean determinants. 
Both of them vanish on the line  $\lambda^S=\frac{1}{2}$ (any  $\lambda^R$), and in addition  $\bar{\Delta}^{(2)}$ vanishes on the line 
 $\lambda^R=\frac{1}{2}$ (any  $\lambda^S$). 
Moreover, there is no domain on the plane $(\alpha_1,\beta_1)$ leading to the  vanishing determinants. There are only two lines 
$\alpha_1=0$ and $\alpha_1=1$ (any $\beta_1$, $\alpha_2$ and $\beta_2$
) yielding the zero determinant 
$\Delta^{(2)}$, while $\Delta^{1}\neq 0$ for any initial 
 parameters $\alpha_i$ and $\beta_i$ (if $\lambda^S\neq \frac{1}{2}$).

\subsubsection{Effect of eigenvector initial parameters $\tilde \Gamma$}
\label{Section:eigenvectors2}

\begin{figure*}
   \epsfig{file=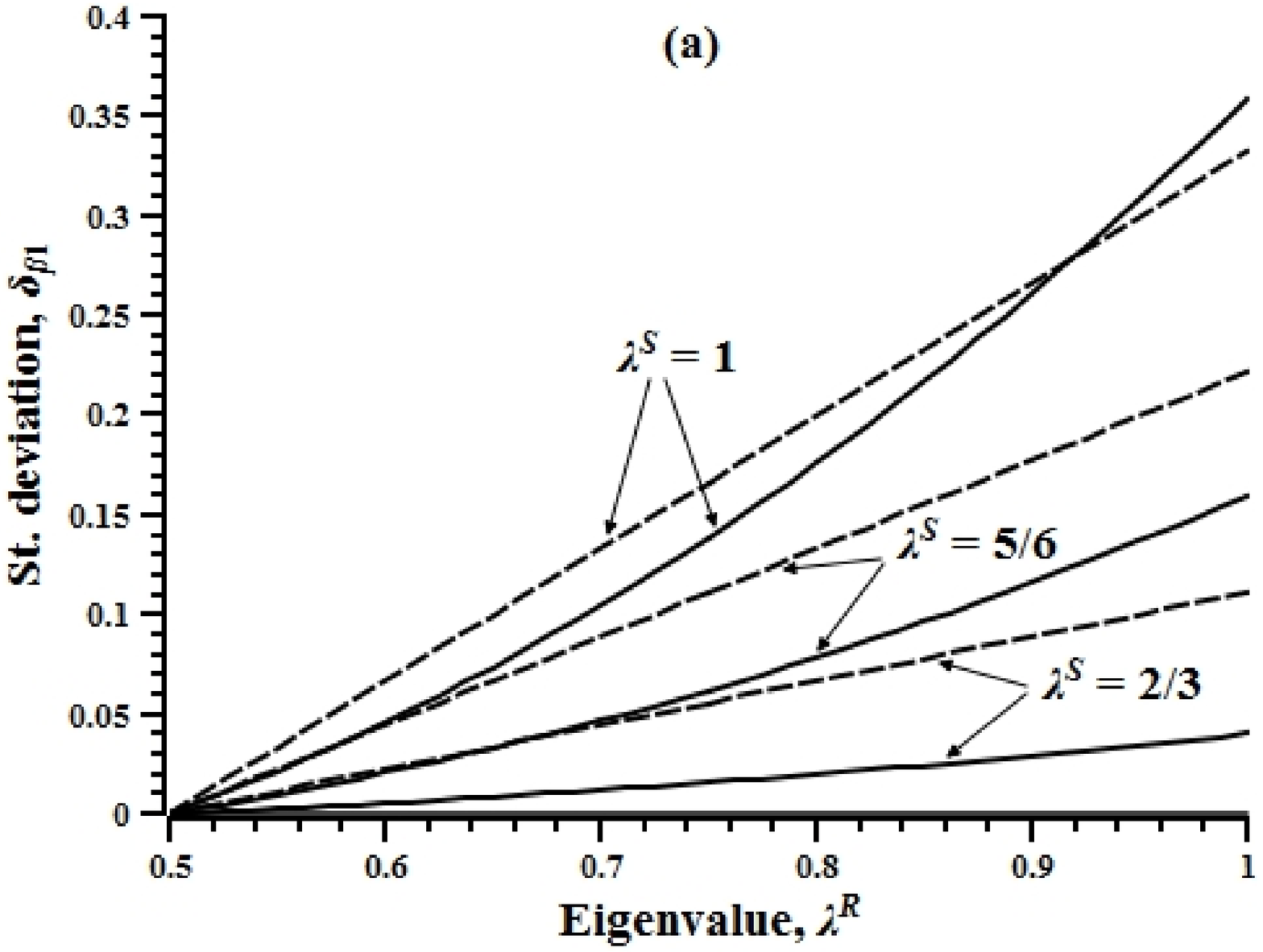,
  scale=0.5
   ,angle=0
}\epsfig{file=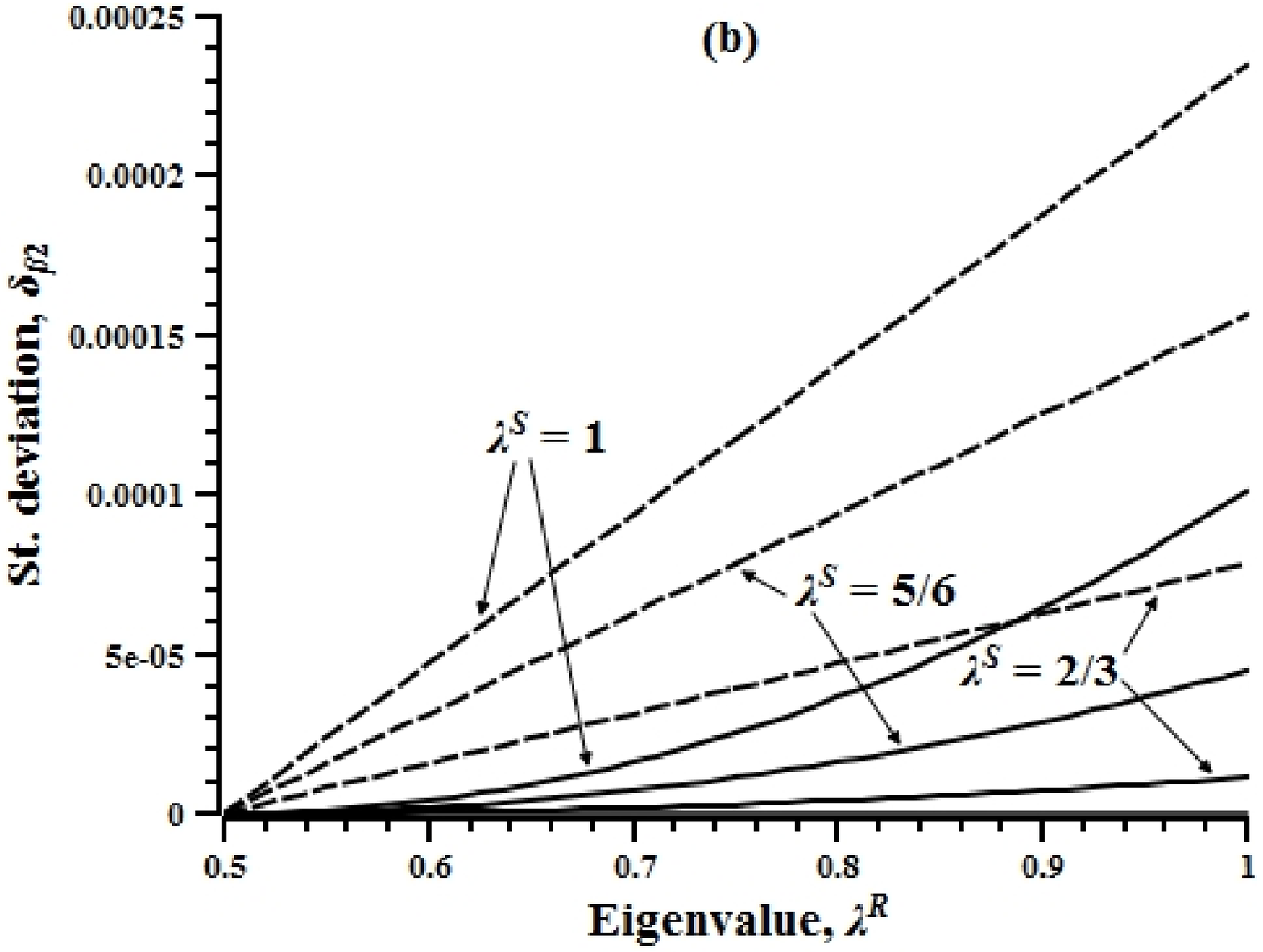,
  scale=0.5
   ,angle=0
}
  \epsfig{file=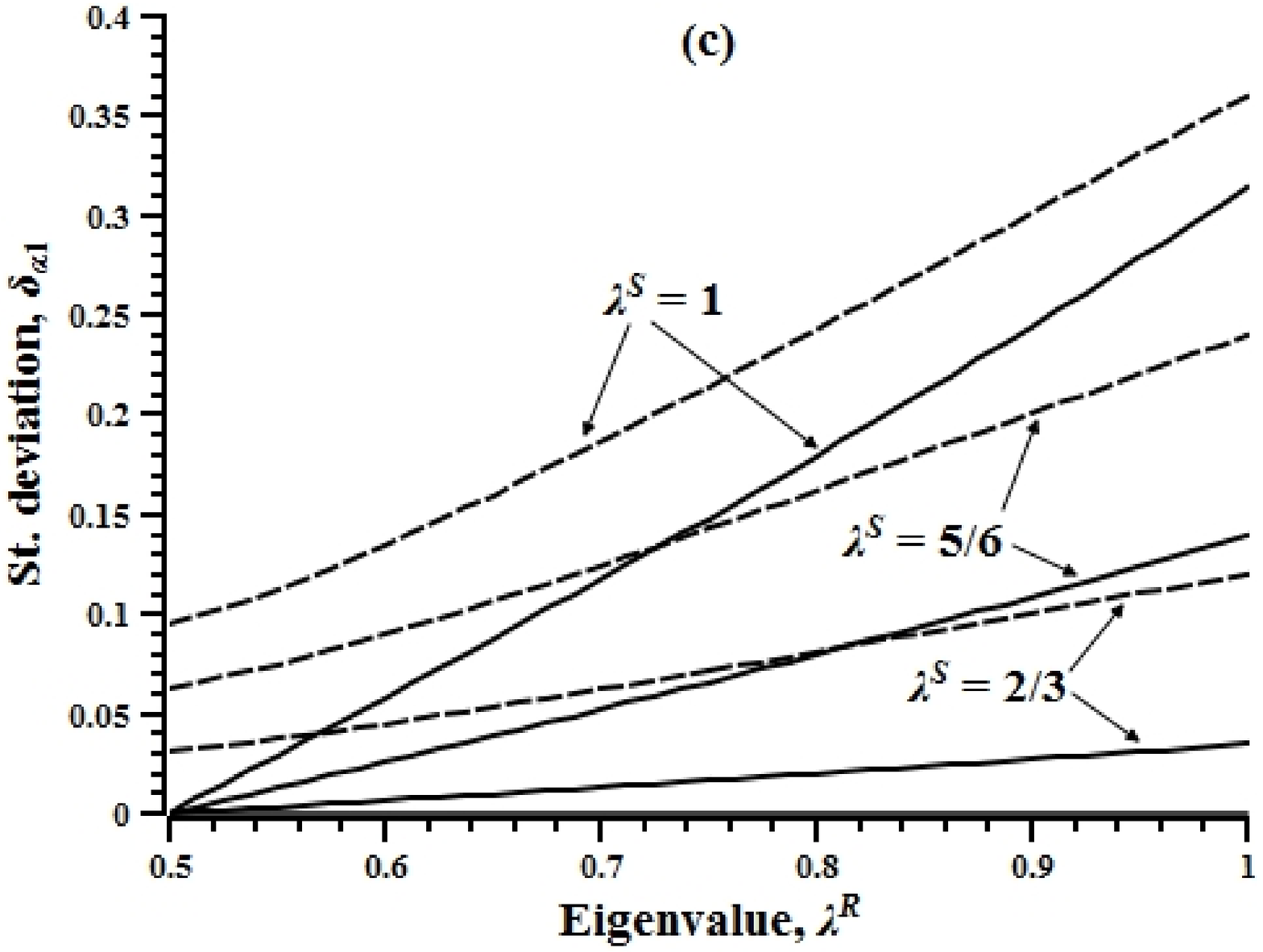,
  scale=0.5
   ,angle=0
}\epsfig{file=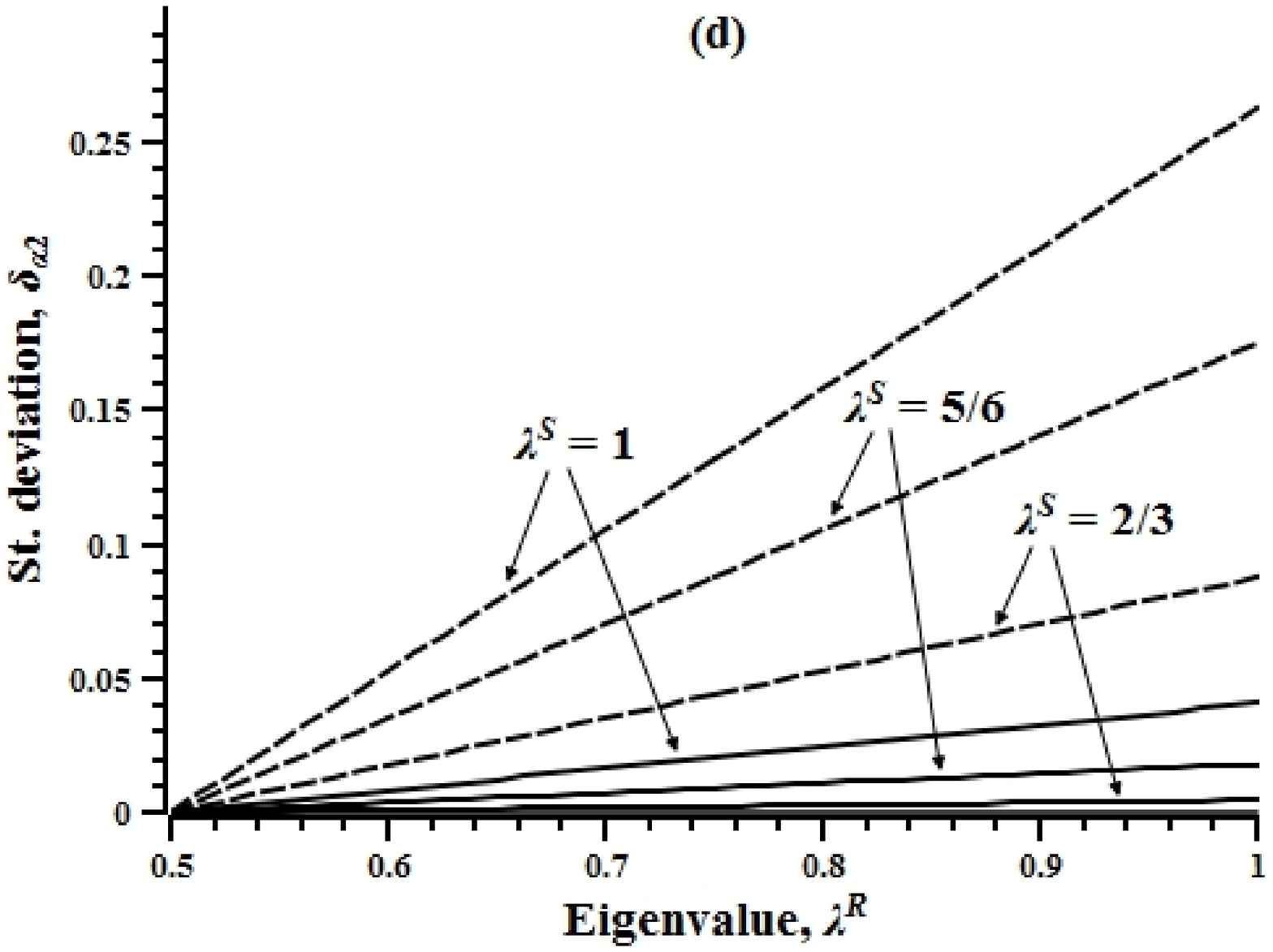,
  scale=0.5
   ,angle=0
}
\caption{ 
The standard deviations  $\delta^{(2)}_{\beta_i}$, $\delta^{(2)}_{\alpha_i}$  (solid lines) and 
 $\delta^{(1)}_{\beta_i}$, $\delta^{(1)}_{\alpha_i}$ (dashed lines), $i=1,2$,
 as
functions of $\lambda^R$ and $\lambda^S$. Each  line corresponds to the particular 
value of $\lambda^S = \frac{1}{2},\frac{2}{3},\frac{5}{6},1$ 
increasing from the bottom to the top of the figure with ${\delta}^{2}_\gamma|_{\lambda^S=\frac{1}{2}}\equiv 
{\delta}^{1}_{\gamma}|_{\lambda^S=\frac{1}{2}}=0$ ($\gamma=\alpha_i,\beta_i$). In addition, 
all the standard deviations vanish at $\lambda^R=\frac{1}{2}$ except $\delta^{(1)}_{\alpha_1}$.
(a) 
$\delta^{(k)}_{\beta_1}$, $k=1,2$, with  their maximal values 
$\delta^{(2)}_{\beta_1}|_{\lambda^S=\lambda^R=1} = 0.3584$, 
$\delta^{(1)}_{\beta_1}|_{\lambda^S=\lambda^R=1} =0.3322$.
(b) 
$\delta^{(k)}_{\beta_2}$, $k=1,2$, 
with their  maximal values
 $\delta^{(2)}_{\beta_2}|_{\lambda^S=\lambda^R=1} = 1.008 \times 10^{-4}$, 
 $\delta^{(1)}_{\beta_2}|_{\lambda^S=\lambda^R=1} = 2.344 \times 10^{-4}$.
(c) 
$\delta^{(k)}_{\alpha_1}$, $k=1,2$, with their maximal values 
$\delta^{(2)}_{\alpha_1}|_{\lambda^S=\lambda^R=1} =0.3137$, 
$\delta^{(1)}_{\alpha_1}|_{\lambda^S=\lambda^R=1} = 0.3601$. In addition, 
$\delta^{(1)}_{\alpha_1}|_{{\lambda^R=\frac{1}{2}}\atop{\lambda^S=1}}=9.477  \times 10^{-2}$.
(d)
$\delta^{(k)}_{\alpha_2}$, $k=1,2$, 
with their maximal values 
$\delta^{(2)}_{\alpha_2}|_{\lambda^S=\lambda^R=1} = 4.067\times 10^{-2}$, 
$\delta^{(1)}_{\alpha_2}|_{\lambda^S=\lambda^R=1} =0.2630$.
}
  \label{Fig:disp1D} 
\end{figure*}

Unlike the concurrence, the informational correlation is not symmetrical with respect
to the replacement $S\leftrightarrow R$, and so do the determinants $\Delta^{(i)}$ ($i=1,2$). 
Therefore, we consider the four 
standard deviations for each determinant $\Delta^{(i)}$:
\begin{eqnarray}\label{deltaD1}
\delta^{\Delta^{(i)}k}_{\beta_i}\equiv \delta^{(k)}_{\beta_i} = \sqrt{\Big\langle \Big(\bar{\Delta}^{(k)} - 
\langle \Delta^{(k)} \rangle_{\tilde\Gamma_{\beta_i}}\Big)^2 \big\rangle_{\beta_i}},
\;\;i,k=1,2,\\\label{deltaD2}
\delta^{\Delta^{(i)}k}_{\alpha_i}\equiv \delta^{(k)}_{\alpha_i} = \sqrt{\Big\langle \Big(\bar{\Delta}^{(k)} - 
\langle \Delta^{(k)} \rangle_{\tilde\Gamma_{\alpha_i}}\Big)^2 \big\rangle_{\alpha_i}},
\;\;i,k=1,2.
\end{eqnarray}
The standard deviations   $\delta^{(k)}_{\beta_i}$
are shown
in Fig.\ref{Fig:disp1D}.
We see similarity in their behavior.  Fig.\ref{Fig:disp1D}a
demonstrates that  determinants
 $\bar{\Delta}^{(i)}$ ($i=1,2$) are sensitive to the parameter $\beta_1$ (strong parameter), 
while, according to Fig.\ref{Fig:disp1D}b, they are non-sensitive to the parameter $\beta_2$
(weak parameter). 
As shown in Fig.\ref{Fig:disp1D}c,d,
both $\alpha_1$ and $\alpha_2$ are strong parameters (i.e., they significantly effect 
the  determinants), contrary to the case of SR-entanglement. 

Remark, that all the standard deviations vanish at $\lambda^R=\frac{1}{2}$, except $\delta^{(1)}_{\alpha_1}$. Therefore, the control parameter 
$\alpha_1$ of the sender  can be considered as the strongest one.

Remember that the informational correlation $E^{SR}$ discussed above depends on the direction of 
the information transfer (from the sender to the  receiver). 
Reversing this direction, we change the strong and weak  eigenvector parameters. Thus, 
 in $E^{SR}$,  the strong parameters are $\alpha_i$, $i=1,2$, and $\beta_1$, while $\beta_2$ is a weak one. 
On the contrary, in  $E^{RS}$, 
the strong parameters are $\beta_i$, $i=1,2$, and $\alpha_1$, while $\alpha_2$ is a weak one.

\section{Comparative analysis of SR-entanglement and  informational correlation}
\label{Section:entinf}

Now we compare the SR-concurrence and determinants  as functions of control parameters. 
As was clearly demonstrated in Sec.\ref{Section:concFeat}, 
if the initial eigenvalues $\lambda^S$ and $\lambda^R$  are below the boundary $B$  (see Fig.\ref{Fig:zero}), 
then  the SR-entanglement 
can not appear during the evolution regardless {of} the values of the control parameters $\alpha_i$ and $\beta_i$. Moreover, 
if  the initial eigenvalues $\lambda^S$ and $\lambda^R$ are above  the boundary $B$, then 
the SR-entanglement can still be zero at  the registration instant unless we take the proper 
values of the control parameters $\alpha_1$ and $\beta_1$. 
In the case of perfect state transfer, the boundary $B$ shrinks to the point 
$(\lambda^R,\lambda^S)=(\frac{1}{2},\frac{1}{2})$.

Meanwhile, the mean  determinants $\bar{\Delta}^{(i)}$ do not vanish both above and below the boundary $B$ (except the particular lines found in Sec. \ref{Section:lambda20}); therefore  
 the eigenvector parameters $\alpha_i$ of the sender's state 
can be  transferred 
to the receiver even if there is no entanglement between these subsystems during the evolution.
All this demonstrates that 
  the SR-entanglement 
is not  responsible for 
 information propagation and  remote state creation because it vanishes in large domain of the control parameters. On the contrary,  
the informational correlation 
is more suitable measure of quantum correlations in this case.

Comparing Fig. \ref{Fig:avrC} with Fig. \ref{Fig:avr} at   $\lambda^R=\frac{1}{2}$,   we see that $\bar C $ doesn't vanish, 
although its value is very small with the maximum
$\bar{C}|_{\lambda^S =1}=1.87\times 10^{-4}$. 
Meanwhile,   $\Delta^{(2)} = 0$ and   $\Delta^{(1)}\neq 0$ 
with the maximal value 
  $\bar{\Delta}^{(1)}|_{\lambda^S =1}=
  0.1929.$
 Thus we can say  that the mean SR-entanglement is non-zero identically if at least one 
 of the eigenvector parameters can be 
 transferred from the sender to the receiver.

 We also see that the  mean values of both   SR-concurrence and determinants are 
 small in the neighborhood of $\lambda^R=\frac{1}{2}$. This means that 
the problem of small concurrence appearing in our model, in certain sense,  
is equivalent to the problem of small determinants 
in linear algebra.

\section{Conclusions}
\label{Section:conclusion}
We study the dependence  of  entanglement and informational correlation 
between the two remote one-qubit  subsystems $S$ and $R$
on the control parameters (which are  the parameters of the sender and receiver initial states). 
The entanglement  is the  well known measure responsible for many advantages 
of quantum information  devices in comparison with their classical counterparts. 
The informational correlation, being based on the parameter exchange between the sender 
and receiver, is closely related to the remote mixed state creation. 
Our basic results are are following.
\begin{enumerate}
\item
There are strong eigenvector control-parameters  which can 
significantly change the quantum correlations. In the case of concurrence,
these are $\alpha_1$ and $\beta_1$. In the case of  informational correlation, 
there are three such parameters: $\alpha_1$, $\alpha_2$ and $\beta_1$ for $E^{SR}$ 
and  $\beta_1$, $\beta_2$ and $\alpha_1$ for $E^{RS}$. 
Other eigenvector control-parameters are weak, they do not essentially effect 
the quantum correlations. These are parameters $\alpha_2$ and $\beta_2$ 
in the case of concurrence. As for the informational correlation, there is  only one weak parameter: 
$\beta_2$ for $E^{SR}$ and $\alpha_2$ for $E^{RS}$. 
\item
The eigenvalues  are most important parameters which 
strongly effect the quantum correlations and, in principle, they might be joined to the above strong
control parameters. However, we keep them in a different group to emphasize the difference between the eigenvector- and eigenvalue control  parameters.
\item
In certain sense, 
there is an equivalence between  the problem of vanishing  
entanglement and the problem of vanishing determinants in linear algebra.
\item
There is a large domain in the  control parameter space mapped into the non-entangled states. 
On the contrary, there is no domain in the control-parameter space leading to zero determinants. The determinants vanish 
only for exceptional values of the control parameters. This fact promotes the informational correlation for a suitable quantity describing the quantum correlations responsible for the state transfer/creation.   
\item
It is remarkable that the weak parameters not only slightly effect on the SR-entanglement and determinants, but have 
a distinguished feature in the problem of remote state creation. Namely, 
according to 
\cite{Z_2014,BZ_2015}, any value of the weak parameter can be created in the receiver's state
using the proper value of the weak parameter of the sender.  
Therefore, the weak parameters can be used for organization of the effective   information transfer 
 without changing the value of SR-entanglement. 
\end{enumerate}
 
 Authors thank Prof. E.B.Fel'dman for useful discussion.
This work is partially supported by the Program of the Presidium of RAS
''Element base of quantum computers''(No. 0089-2015-0220) and 
by the Russian Foundation for Basic Research, grant No.15-07-07928.

\section{Appendix}
\label{Section:appendix}

\subsection{Permanent characteristics of communication line}
\label{Section:appendixA}
Writing $\rho^{SR}$ (\ref{RhoSR})   in components, we have
\begin{eqnarray}\label{rhoR}
\rho^{SR}_{i_1 i_N; j_1 j_N} = T_{i_1 i_N l_1 l_N;j_1 j_N k_1k_N}(\rho^S_0)_{l_1; k_1}(\rho^R_0)_{l_N; k_N}.
\end{eqnarray}
Here all the indexes take two values $0$ and  $1$, the parameters $T_{i_1 i_N l_1 l_N;j_1 j_N k_1k_N}$
in this formula depend on the Hamiltonian as follows:
\begin{eqnarray}\label{T0}
 T_{i_1 i_N l_1 l_N;j_1 j_N k_1k_N} = \sum_{i_{TL},l_{TL},k_{TL}}
 V_{i_1 i_{TL} i_N ;l_1 l_{TL} l_N } \rho^{TL}_{l_{TL};k_{TL}}
 V^+_{ k_1 k_{TL} k_N; j_1 i_{TL} j_N},
\end{eqnarray}
where the indexes with the subscript $TL$ are the vector indexes of $(N-2)$ scalar binary  indexes,
for instance: 
$i_{TL} =\{ i_2\dots i_{N-1}\}$. We refer to these parameters as $T$-parameters. 
In formulas (\ref{rhoR}) and (\ref{T0}), we write the components of both the density matrices 
and the operator $V$, where both rows and columns are enumerated by 
the vector subscripts consisting of the binary  indexes. For instance,
\begin{eqnarray}\nonumber
\rho^{SR}_{i_1 i_N; j_1 j_N} \;\; {\mbox{is the element at the intersection of  row}} \;\; \{i_1 i_N\}
\;\; {\mbox{and  column}} \;\; \{j_1 j_N\},
\end{eqnarray}
and similar for the components of the operator $V$.

If the transmission line is in ground state (\ref{TL}),
 then 
the expression for the  $T$-parameters  is simpler:
\begin{eqnarray}\label{T}
 T_{i_1 i_N l_1 l_N;j_1 j_N k_1k_N} = \sum_{i_{TL}}
V_{i_1 i_{TL} i_N ;l_1 0_{TL} l_N }  V^+_{ k_1 0_{TL} k_N; j_1 i_{TL} j_N},
\end{eqnarray}
where $0_{TL} =(\underbrace{0,\dots,0}_{N-2})$.
The number of $T$-parameters is independent on the length of a transmission line 
and is completely defied by the 
dimensionality of the sender and receiver. 

The $T$-parameters have two  obvious symmetries. 
The first one  follows from the Hermitian property of the density matrix (\ref{rhoR}),
$(\rho^{SR})^+=\rho^{SR}$:
\begin{eqnarray} \label{sym1} 
T_{i_1 i_N l_1 l_N; j_1j_N k_1 k_N} =T_{j_1j_N k_1 k_N;i_1 i_N l_1 l_N}^*\;\;
\Rightarrow \;\;
{\mbox{Im}} \; T_{j_1j_N k_1 k_N;j_1 j_N k_1 k_N} =0.
\end{eqnarray}
The second symmetry follows from the fact that these parameters must be 
symmetrical with respect to the exchange  $S\leftrightarrow R$:
\begin{eqnarray}\label{sym2} 
T_{i_1 i_N l_1 l_N; j_1j_N k_1 k_N} = T_{i_N i_1 l_N l_1; j_Nj_1 k_N k_1}.
\end{eqnarray}
Finally, the set of  $T$-parameters equals zero as a consequence of the fact that the 
Hamiltonian commutes with  the $z$-projection of the total momentum $I_z$;
therefore the nonzero   elements $V_{I,J}$ of the evolution operator  are those, 
whose  $N$-dimensional vector indexes $I$ and $J$ have equal number of units. Consequently (if the transmission line $TL$ is in ground state 
(\ref{TL}) initially),
\begin{eqnarray}\label{T00}
T_{i_1 i_N l_1 l_N; j_1j_N k_1 k_N} =0 \;\;{\mbox{if}}\;\;
\left\{
\begin{array}{l}
i_1+i_N > l_1+l_N \cr
j_1+j_N > k_1+k_N \cr
i_1+i_N < l_1+l_N\;\;{\mbox{and}}\;\; i_1+i_N -(j_1+j_N)\neq l_1+l_N-(k_1+k_N)
\end{array}
\right..
\end{eqnarray}
In other words, the following  $T$-parameters are nonzero:
\begin{eqnarray}
&&T_{i_1 i_N l_1 l_N; j_1j_N k_1 k_N} \neq0 \;\;{\mbox{if}}\;\;\\\nonumber
&&
(i_1+i_N \le l_1+l_N) \wedge (j_1+j_N \le k_1+k_N)\wedge 
(i_1+i_N -j_1-j_N = l_1+l_N-k_1-k_N).
\end{eqnarray}
The  $T$-parameters  are permanent
characteristics of the communication line which do not change during its operation.

\subsection{Explicit form for  elements of  receiver's density matrix}
\label{Section:appendixB}
We  obtain the element of the  receiver's density matrix $\rho^R(t)$ calculating the trace of the matrix 
$\rho^{SR}$  (\ref{rhoR}) over the sender's node:
\begin{eqnarray}\label{rhoR2}
\rho^{R}_{i_N;  j_N}=\sum_{i_1,j_1} \rho^{SR}_{i_1 i_N; j_1 j_N} = 
T_{ i_N l_1;j_N k_1}(\rho^S_0)_{l_1; k_1} ,
\end{eqnarray}
where
\begin{eqnarray}\label{TT2}
T_{ i_N l_1 ; j_N k_1}=\sum_{l_N,k_N,i_1} T_{i_1 i_N l_1 l_N;i_1 j_N k_1k_N}(\rho^R_0)_{l_N; k_N},
\end{eqnarray}
and $T_{ i_N l_1 ; j_N k_1}$ satisfies the symmetry following from symmetry (\ref{sym1}):
\begin{eqnarray} \label{sym12} 
T_{ i_N l_1 ;j_N k_1 } =T_{j_N k_1;i_N l_1}^*\;\;
\Rightarrow \;\;
{\mbox{Im}} \; T_{i_N l_1 ;i_N l_1 } =0.
\end{eqnarray}
In result, the independent elements of $\rho^R$ read as follows:
\begin{eqnarray}\label{y0}
\rho^{R}_{1;1} = T_{1 0;1 0} +(T_{1 1;1 1}-T_{1 0;1 0}) x_1 +(T_{1 0;1 1} + T_{1 1;1 0}) x_2 
+i (T_{1 0;1 1} - T_{1 1;1 0}) x_3 ,\\
\rho^{R}_{0;1} = T_{0 0;1 0} +(T_{0 1;1 1}-T_{0 0;1 0}) x_1 +(T_{0 0;1 1} + T_{0 1;1 0}) x_2 
+i (T_{0 0;1 1} - T_{0 1;1 0}) x_3,
\end{eqnarray}
which is a system of linear algebraic equations allowing us to determine the initial parameters $x_i$
knowing the registered density matrix  of the receiver's state. 
We can conveniently rewrite  system (\ref{y0})
separating the real and imaginary parts to get three independent real equations (\ref{y}).

\subsection{Some properties of determinants}
\label{Section:appendixC}
\label{Section:propdet}
The both determinants  $\Delta^{(1)}$ and $\Delta^{(2)}$ depend on the
parameters of the initial states of the sender and receiver: 
$\lambda^S$, $\lambda^R$, $\alpha_i$, $\beta_i$, $i=1,2$. 
But this dependence is partially separated, which has been already used in eqs.(\ref{normD}):
expressions  $\left|\frac{\partial(y_i,y_j)}{\partial(x_n,x_m)}\right|$ 
and $\left|\frac{\partial y_i}{\partial x_n}\right|$ 
in, respectively,  eqs.(\ref{Delta2})
 and (\ref{Delta1}) depend
 on $\lambda^R$, $\beta_i$,
 $i=1,2$, while 
expressions $\left|
\frac{\partial(x_n ,x_m)}{\partial(\alpha_1,\alpha_2)}\right|$
and $\left(\left|\frac{\partial x_n }{\partial \alpha_1} \right| +
\left|\frac{\partial x_n }{\partial \alpha_2} \right|\right)$ 
in, respectively, eqs.(\ref{Delta2})
 and (\ref{Delta1})
depend on  $\lambda^S$, $\alpha_i$, $i=1,2$. All this immediately follows from the definitions of $x_i$ (\ref{x}) and  elements of $\rho^R$ (\ref{y}).

Notice that each term in definitions (\ref{Delta1}) and (\ref{Delta2}) 
is the independent determinant condition
for solvability of  system (\ref{y}) for, respectively,
two parameters $\alpha_i$, $i=1,2$,  or one of them. In other words, if there are $k$ nonzero terms 
in these formulas, then we can find parameters $\alpha_i$ ($i=1,2$)  in $k$ different ways. 
In principle, if each term is small in eq.(\ref{Delta2}) (or (\ref{Delta1})), then 
the parameters $\alpha_1$ and $\alpha_2$ (or one of them)   can be found from  system (\ref{y})  with restricted accuracy. 
However, if there are $k$  small but nonzero terms in (\ref{Delta1}) (or  (\ref{Delta2})), then 
the accuracy can be improved by calculating the transferred  parameters  $k$ times and comparing the results. 
For this reason we do not divide the sums in both formulas (\ref{Delta1}) and  (\ref{Delta2}) 
by the number of terms in them.

\subsection{Choice of time instant for state registration}
\label{Section:timeinst}
\label{Section:appendix0}
Now we show that $C$ and
the determinants  $\Delta^{(i)}$  averaged over the 
initial conditions 
are maximal at the  time instant of the maximum of $\langle \bar{P}\rangle_{\lambda^S,\lambda^R}(t) = \bar{P}(t)$, where
\begin{eqnarray}\label{barP}
\bar P(t)&\equiv& \langle P\rangle_{\tilde\Gamma}=
\frac{1}{P_0}\left\langle \rho^{(SR)}_{01;01} (t)+ \rho^{(SR)}_{10;10} (t)+ 
\rho^{(SR)}_{11;11}(t)\right\rangle_{\tilde \Gamma}=\\\nonumber
&&
\frac{2}{3}\left(
T_{0110;0110}(t)  +T_{1010;1010}(t)+ T_{1011;1011}(t) + \frac{1}{2}T_{1111;1111}(t)
\right).
\end{eqnarray}
Here $P_0=\frac{3}{4}$ is the normalization fixed by the requirement 
$\bar P|_{t=0}=1$ and we take into account that  $\bar P$ doesn't depend on the initial 
eigenvalues $\lambda^S$ and $\lambda^R$.
The function 
$P$ can be viewed  as a  probability  of registration of the excitation 
at the nodes of the subsystem $SR$.  
{ The numerical calculations show that } its maximum coincides with the maximum of fidelity 
of a one-qubit pure 
state transfer:
\begin{eqnarray}
\bar P^R(t)=
\rho^{(SR)}_{01;01}(t)|_{ \Gamma=\{0,1,0,0,0,0\}}=
T_{0110;0110}(t) 
.
\end{eqnarray}
This fact simplifies our calculations.

\begin{figure*}
   \epsfig{file=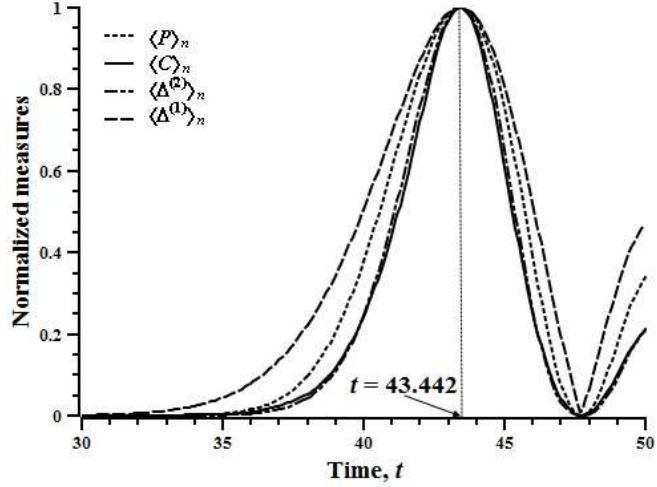,
  scale=0.5 ,angle=0
}
\caption{The time-dependence of the normalized  mean probability
$\langle\bar{P}\rangle_{n}$
(dotted line),  mean SR-concurrence 
$\langle \bar{C} \rangle_{n}$ (solid line), mean
determinants
$\langle \bar{\Delta}^{(2)} \rangle_{n}$ (dash-dotted line) 
and 
$\langle \bar{\Delta}^{(1)} \rangle_{n}$ (dashed line) defined in eq.(\ref{PCDnorm}) with normalizations given in (\ref{norm}).
All four curves have the maximum at the same time instant
$t=43.442$ (we use the values of  $T$-parameters found in Appendix, Sec.\ref{Section:appendixD}). 
} 
  \label{Fig:maxPDeltaC} 
\end{figure*}

The time-dependences of the functions
 $\langle\bar {P}\rangle_{\lambda^S,\lambda^R}\equiv \bar {P}$, 
$\langle \bar{\Delta}^{(i)} \rangle_{\lambda^S,\lambda^R}$ and 
$\langle \bar C \rangle_{\lambda^S,\lambda^R}$  are shown in 
Fig. \ref{Fig:maxPDeltaC}
for the chain of $N=40$ nodes
(for convenience, we normalize them by their maxima over the considered long enough interval, $0\le t \le 50$ , i.e., we show 
the ratios
 \begin{eqnarray}\label{PCDnorm}
\langle {P} \rangle_n= \frac{\langle \bar{P} \rangle_{\lambda^S,\lambda^R}}{
\langle \bar{P} \rangle_{\lambda^S,\lambda^R}^{max}},\;\;
\langle {C} \rangle_n= \frac{\langle \bar{C} \rangle_{\lambda^S,\lambda^R}}{
\langle \bar{C} \rangle_{\lambda^S,\lambda^R}^{max}},\;\;
\langle {\Delta}^{(i)} \rangle_n=\frac{\langle \bar{\Delta}^{(i)} \rangle_{\lambda^S,\lambda^R}}{
\langle \bar{\Delta}^{(i)} \rangle_{\lambda^S,\lambda^R}^{max}},
\end{eqnarray}
where 
\begin{eqnarray}\label{norm}
&&
\langle \bar{P} \rangle_{\lambda^S,\lambda^R}^{max}=\langle \bar{P} \rangle_{\lambda^S,\lambda^R}|_{t=43.442}=0.5476,\;\;
\langle \bar{C} \rangle_{\lambda^S,\lambda^R}^{max}=\langle \bar{C} \rangle_{\lambda^S,\lambda^R}|_{t=43.442}=9.584\times 10^{-3},\\\nonumber
&&
\langle \bar{\Delta}^{(2)} \rangle_{\lambda^S,\lambda^R}^{max}=\langle \bar{\Delta}^{(2)} \rangle_{\lambda^S,\lambda^R}|_{t=43.442}=9.846\times 10^{-2},\;\;
\langle \bar{\Delta}^{(1)} \rangle_{\lambda^S,\lambda^R}^{max}=\langle \bar{\Delta}^{(1)} \rangle_{\lambda^S,\lambda^R}|_{t=43.442}=0.2765.
\end{eqnarray}
We see that the time instant of  the maxima   is the same for  all four functions  and equals $t=43.442$.
Namely this optimized time instant is taken for our calculations.

\subsection{Numerical values of  $T$-parameters for $N=40$ at  optimized time instant.}
\label{Section:appendixD}
For the case $N=40$, we have calculated  the $T$-parameters at the optimized time instant $t=43.442$ found in Sec.\ref{Section:timeinst}. 
 Similar to \cite{SZ_2016}, the $T$-parameters
 can be separated into three families by  their absolute values.
We give the list of these families up to symmetries (\ref{sym1},\ref{sym2}).

{\bf 1st family:} There are two different parameters with the absolute values  gapped in the interval $[6.817\times 10^{-1},1]$:
\begin{eqnarray}
T_{0000;0000}=1, T_{0000;0110}=-6.817 i \times 10^{-1}.
\end{eqnarray}

{\bf 2nd family:} There are 8 different  parameters with the absolute values  gapped in the interval 
$[2.160\times 10^{-1} ,5.353\times 10^{-1} ]$:
\begin{eqnarray}
&&
T_{0001;0001}=5.352 \times  10^{-1} ,\;\;
 T_{0 0 1 1;0 0 1 1}= 2.865  \times 10^{-1},\;\;\\\nonumber
 &&
 T_{0 0 0 1;0 1 1 1}= 3.649 i  \times 10^{-1},\;\;
 T_{0 0 0 0;1 1 1 1}= 4.648  \times 10^{-1},\;\;\\\nonumber
 &&
 T_{0 1 1 0;0 1 1 0}= 4.648  \times 10^{-1},\;\;
 T_{0 1 1 1;0 1 1 1}= 2.488 \times  10^{-1} ,\;\;\\\nonumber
 &&
 T_{0 1 1 0;1 1 1 1}= 3.169 i  \times 10^{-1},\;\;
 T_{1 1 1 1;1 1 1 1}= 2.160  \times 10^{-1}.
\end{eqnarray}

{\bf 3rd family:} There are 5 different parameters with the absolute values gapped in the interval $[0,5.396 \times 10^{-3}]$:
\begin{eqnarray}
&&
T_{0 0 0 0;0 1 0 1}= -5.395  \times 10^{-3}, \;\;
 T_{0 0 1 0;0 1 1 1}= -2.888  \times 10^{-3},\\\nonumber
 &&
 T_{0 1 0 1;0 1 0 1}= 2.911  \times 10^{-5},\;\;
 T_{0 1 0 1;0 1 1 0}= 3.678 i  \times 10^{-3},\\\nonumber
 &&
 T_{0 1 0 1;1 1 1 1}= -2.508  \times 10^{-3}.
\end{eqnarray}
Notice that the parameter $T_{0001;0010}$ vanishes only due to the nearest-neighbor interaction
model and/or even $N$. It becomes non-vanishing 
if at least one of these conditions is destroyed.

We see that there are certain gaps between the neighboring families, which is most significant  ($\sim 10^2$)
between the 2nd and the 3rd 
families. 
In addition,  the parameters from the 3rd family are  smallest ones. Similar to ref.\cite{SZ_2016}, 
this difference in absolute values of the $T$-parameters is due to the symmetries of transitions among the different nodes of the chain. 
The obtained values of the $T$-parameters   are used in Sec.\ref{Section:lambda20}.


\begin{thebibliography}{99}

 
\bibitem{HW}
S.Hill and W.K.Wootters, Phys. Rev. Lett. {\bf    78}, 5022 (1997)


\bibitem{Wootters}
 W.K. Wootters,
Phys. Rev. Lett. {\bf   80},
2245 (1998)

 

\bibitem{BDFMRSSW}
 Bennett, C.H., DiVincenzo, D.P., Fuchs, C.A., Mor, T., Rains, E., Shor, P.W., Smolin, J.A., Wootters,
W.K.,
Phys. Rev. A {\bf 59}, 1070 (1999)

\bibitem{HHHOSS}
Horodecki, M., Horodecki, P., Horodecki, R., Oppenheim, J., Sen, A., Sen, U., Synak-Radtke, B.,
Phys. Rev. A {\bf 71}, 062307 (2005)

\bibitem{NC}
Niset, J., Cerf, N.J.,
Phys. Rev. A {\bf 74}, 052103 (2006)

 
\bibitem{Meyer}
Meyer, D.A.
Phys. Rev. Lett. {\bf 85}, 2014 (2000)

 
 
\bibitem{DSC}
A.Datta, A.Shaji,  C.M.Caves,
Phys.Rev.Lett. {\bf 100}, 050502 (2008)

 
 
\bibitem{DFC}
Datta, A., Flammia, S.T., Caves, C.M.,
Phys. Rev. A {\bf 72}, 042316 (2005)

\bibitem{DV}
Datta, A., Vidal, G.,
Phys. Rev. A {\bf 75}, 042310 (2007)

 
 
 
\bibitem{LBAW}
 B.P.Lanyon, M.Barbieri, M.P.Almeida, A.G.White,
 Phys.Rev.Lett. {\bf 101}, 200501 (2008)

 
\bibitem{Z0}
W. H. Zurek, Ann. Phys.(Leipzig), {\bf 9}, 855 (2000)

\bibitem{HV}
L.Henderson and V.Vedral J.Phys.A:Math.Gen. {\bf 34}, 6899 (2001)

\bibitem{OZ}
H.Ollivier, W.H.Zurek,
Phys. Rev. Lett. {\bf 88}, 017901 (2001)


\bibitem{Z}
W. H. Zurek, Rev. Mod. Phys. {\bf 75}, 715 (2003)
 
 
 

\bibitem{ZZHE}
M.Zukowski, A.Zeilinger, M.A.Horne, A.K.Ekert, Phys. Rev.
Lett. {\bf 71}, 4287 (1993)
 
\bibitem{BPMEWZ}
D.Bouwmeester, J.-W. Pan, K.Mattle, M.Eibl, H.Weinfurter, and  A. Zeilinger, 
Nature {\bf 390}, 575 (1997)

\bibitem{BBMHP}
D. Boschi,  S. Branca,  F. De Martini, L. Hardy,  and S. Popescu,
Phys. Rev. Lett. {\bf 80}, 1121 (1998)

 
 
 
 
 
 
\bibitem{PBGWK2}
N.A.Peters, J.T.Barreiro,  M.E.Goggin, T.-C.Wei,  and P.G.Kwiat, Phys.Rev.Lett. {\bf   94}, 
150502 (2005) 

\bibitem{PBGWK}
N.A.Peters, J.T.Barreiro, M.E.Goggin, T.-C.Wei, and P.G.Kwiat in {\it Quantum
Communications and Quantum Imaging III}, ed. R.E.Meyers,
Ya.Shih, Proc. of SPIE {\bf   5893} (SPIE, Bellingham, WA, 2005) 

\bibitem{DLMRKBPVZBW}
B.Dakic, Ya.O.Lipp, X.Ma, M.Ringbauer, S.Kropatschek,
S.Barz, T.Paterek, V.Vedral, A.Zeilinger, C.Brukner, and P.Walther, 
Nat. Phys. {\bf   8}, 666 (2012). 

\bibitem{XLYG}
G.Y. Xiang, J.Li, B.Yu, and G.C.Guo
Phys. Rev. A {\bf   72}, 012315  (2005)

\bibitem{LH}
  L.L.Liu, T. Hwang,
 Quantum Inf. Process. 
 {\bf 13}, 1639 (2014)




 
\bibitem{Bose}
S. Bose, Phys. Rev. Lett. {\bf   91}, 207901 (2003)


 
\bibitem{CDEL}
 M.Christandl, N.Datta, A.Ekert and A.J.Landahl, Phys.Rev.Lett. {\bf   92}, 187902 (2004)

\bibitem{ACDE}
 C.Albanese, M.Christandl, N.Datta and A.Ekert, Phys.Rev.Lett. {\bf   93}, 230502 (2004)

\bibitem{KS}
 P.Karbach and J.Stolze, Phys.Rev.A {\bf   72}, 030301(R) (2005)
 
\bibitem{GKMT}
 G.Gualdi, V.Kostak, I.Marzoli and P.Tombesi, Phys.Rev. A {\bf   78}, 022325 (2008)


 
\bibitem{WLKGGB}
A.W\'ojcik, T.Luczak, P.Kurzy\'nski, A.Grudka, T.Gdala, and M.Bednarska
Phys. Rev. A {\bf   72}, 034303 (2005)
 
\bibitem{NJ}
G.M.Nikolopoulos and I.Jex, eds., {\it Quantum State Transfer and Network Engineering}, Series in
Quantum Science and Technology, Springer, Berlin Heidelberg (2014)



 \bibitem{SAOZ}
 J.Stolze, G. A. \'Alvarez,
O. Osenda, A. Zwick in
{\it Quantum State Transfer and Network Engineering.
Quantum Science and Technology},
ed. by  G.M.Nikolopoulos and I.Jex, Springer Berlin Heidelberg, Berlin, p.149  (2014) 


 
\bibitem{CS}
 B. Chen and Zh. Song, Sci. China-Phys., Mech.Astron. 53, 1266
(2010).
 
 \bibitem{BOWB}
C. A. Bishop, Yo.-Ch. Ou, Zh.-M. Wang, and M. S. Byrd,
Phys. Rev. A {\bf 81}, 042313 (2010)


 \bibitem{ABCVV}
T. J. G. Apollaro, L. Banchi, A. Cuccoli, R. Vaia, and P. Verrucchi,
Phys. Rev. A {\bf 85}, 052319 (2012)
 
 
 \bibitem{Z_inf0}
  A.I.Zenchuk,   J. Phys. A: Math. Theor.  {\bf 45} (2012) 115306
 
 
\bibitem{Z_inf}
A.I.Zenchuk, Quant. Inf. Proc. {\bf 13}, 2667-2711 (2014)

 
\bibitem{FY}
E. B. Fel'dman , M. A. Yurishchev,
JETP Letters {\bf 90}, 70 (2009)

\bibitem{Y}
M. A. Yurishchev, 
JETP {\bf 111}, 525 (2010)
 
 


\bibitem{SZ_2016}
J.Stolze, A.I.Zenchuk, to appear in Quant.Inf.Proc. {\bf 15}, 3347-3366 (2016)
 arXiv:1512.04309


\bibitem{Z_2014}
A.I.Zenchuk, 
Phys. Rev. A {\bf 90}, 052302(13) (2014) 

 
\bibitem{BZ_2015}
G. A. Bochkin and A. I. Zenchuk, 
Phys.Rev.A 91, 062326(11) (2015)
 
 
 

\end{thebibliography}
\end{document}